\pdfoutput=1
\documentclass[prd,onecolumn,notitlepage,eqsecnum,nofootinbib,floatfix]{revtex4-1}
\usepackage{amssymb}
\usepackage{amsmath}
\usepackage{amsfonts} 
\usepackage{bm}
\usepackage{graphicx}
\DeclareFontFamily{OT1}{rsfs}{} 
\DeclareFontShape{OT1}{rsfs}{m}{n}{<-7> rsfs5 
   <7-10> rsfs7 <10-> rsfs10}{}   
\DeclareMathAlphabet{\scr}{OT1}{rsfs}{m}{n} 
\DeclareSymbolFont{EulerScript}{U}{eus}{m}{n}
\SetSymbolFont{EulerScript}{bold}{U}{eus}{b}{n}
\DeclareSymbolFontAlphabet\scrpt{EulerScript}
\newcommand{\LL}{{\scr L}} 
\newcommand{\M}{{\scr M}} 
\newcommand{\E}{{\cal E}} 
\newcommand{\B}{{\cal B}} 
\newcommand{\K}{{\cal K}} 
\newcommand{\Bq}{{\cal B}^{\scriptstyle \sf q}} 
\newcommand{\Bhatq}{\hat{\cal B}^{\scriptstyle \sf q}} 
\newcommand{\Kd}{{\cal K}^{\scriptstyle \sf d}} 
\newcommand{\Ko}{{\cal K}^{\scriptstyle \sf o}} 
\newcommand{\fd}{{\lambda}^{\scriptstyle \sf d}} 
\newcommand{\fo}{{\lambda}^{\scriptstyle \sf o}} 
\newcommand{\pd}{{p}^{\scriptstyle \sf d}} 
\newcommand{\po}{{p}^{\scriptstyle \sf o}} 
\newcommand{\Ud}{{U}^{\scriptstyle \sf d}} 
\newcommand{\Uo}{{U}^{\scriptstyle \sf o}} 
\newcommand{\zetad}{{\zeta}^{\scriptstyle \sf d}} 
\newcommand{\zetao}{{\zeta}^{\scriptstyle \sf o}} 
\newcommand{\zetahatq}{{\hat{\zeta}}^{\scriptstyle \sf q}} 
\newcommand{\wwd}{{w}^{\scriptstyle \sf d}} 
\newcommand{\wwo}{{w}^{\scriptstyle \sf o}} 
\newcommand{\whatq}{{\hat{w}}^{\scriptstyle \sf q}}  
\newcommand{\yd}{{\sf y}^{\scriptstyle \sf d}} 
\newcommand{\yo}{{\sf y}^{\scriptstyle \sf o}} 
\newcommand{\zd}{{\sf z}^{\scriptstyle \sf d}} 
\newcommand{\ppd}{{\sf p}^{\scriptstyle \sf d}} 
\newcommand{\ppo}{{\sf p}^{\scriptstyle \sf o}} 
\newcommand{\UUd}{{\sf U}^{\scriptstyle \sf d}} 
\newcommand{\UUo}{{\sf U}^{\scriptstyle \sf o}} 
\newcommand{\m}{{\sf m}} 
 
\newcommand{\KKo}{\mathfrak{K}^{\scriptstyle \sf o}} 
\newcommand{\kko}{\mathfrak{k}^{\scriptstyle \sf o}} 
\allowdisplaybreaks
\begin{document}
\title{Gravitomagnetic tidal currents in rotating neutron stars} 
\author{Eric Poisson and Jean Dou\c{c}ot}  
\affiliation{Department of Physics, University of Guelph, Guelph,
  Ontario, N1G 2W1, Canada} 
\date{January 25, 2017} 
\begin{abstract} 
It was recently revealed that a rotating compact body responds
dynamically when it is subjected to a gravitomagnetic tidal field,
even when this field is idealized as time-independent. The dynamical
response is characterized by time-changing internal currents, and it
was suspected to originate from zero-frequency $g$-modes and $r$-modes 
driven by the tidal forces. In this paper we provide additional
insights into the phenomenon by examining the tidal response of a
rotating body within the framework of post-Newtonian gravity. This
approach allows us to develop an intuitive picture for the phenomenon,
which relies on the close analogy between post-Newtonian gravity and
Maxwell's theory of electromagnetism. In this picture, the coupling
between the gravitomagnetic tidal field and the body's rotational
velocity is naturally expected to produce an unbalanced Lorentz-like
force within the body, and it is this force that is responsible for
the tidal currents. The simplicity of the fluid equations in the
post-Newtonian setting allows us to provide a complete description of
the zero-frequency modes and demonstrate their precise role in the 
establishment of the tidal currents. We estimate the amplitude of
these currents, and find that for neutron-star binaries of relevance
to LIGO, the scale of the velocity perturbation is measured in
kilometers per second when the rotation period is comparable to 100
milliseconds. This estimate indicates that the tidal currents may
have a significant impact on the physics of neutron stars near merger.   
\end{abstract} 
\maketitle

\section{Introduction} 
\label{sec:intro} 

The tidal interaction between neutron stars in a close binary system 
has recently been the subject of intense investigation, following the
remarkable observation \cite{flanagan-hinderer:08, hinderer:08} that
the tidal deformation of each body could have a measurable impact on 
the emitted gravitational waves. The effect depends on the tidal
polarizability of each neutron star, and a large effort was
devoted to the computation of relativistic Love numbers 
\cite{damour-nagar:09, binnington-poisson:09, landry-poisson:14} 
for realistic models of neutron stars, and to ascertain the importance
of the tidal deformation on the gravitational-wave signal
\cite{hinderer-etal:10,  baiotti-etal:10, 
  baiotti-etal:11, vines-flanagan-hinderer:11, pannarale-etal:11,
  lackey-etal:12, damour-nagar-villain:12,  bini-damour-faye:12, read-etal:13,
  vines-flanagan:13, maselli-gualtieri-ferrari:13,
  chakrabarti-delsate-steinhoff:13a, chakrabarti-delsate-steinhoff:13b,
  lackey-etal:14, dolan-etal:14, bini-damour:14, favata:14,
  yagi-yunes:14, delsate:15, pani-etal:15a, pani-etal:15b}. While 
this work was restricted to the regime of static tides (or adiabatic
tides), in which the external, orbital time scale is long
compared with the internal, hydrodynamical time scale of the neutron
star, an extension to the regime of dynamical tides was recently
developed \cite{ferrari-gualtieri-maselli:12, maselli-etal:12,
  maselli-etal:13, steinhoff-etal:16}, following the pioneering work
of Flanagan and Hinderer \cite{flanagan-hinderer:08}. The dynamical
aspects of the tidal interaction, during which the body's internal
fluid modes are driven by the external tidal forces, were shown to be 
significant for binaries with mass ratios up to approximately 3, when
they implicate stiff neutron stars with large radii \cite{steinhoff-etal:16}.  

An unexpected aspect of the tidal dynamics of compact binaries was
recently revealed by Landry and Poisson
\cite{landry-poisson:15c}. These authors demonstrated that a rotating
compact body responds dynamically when it is subjected to
a {\it gravitomagnetic} tidal field --- the inhomogeneous piece of the 
gravitational field produced by the mass current associated with the 
orbital motion of the companion body. Most strikingly, the phenomenon
was revealed in the idealization in which the gravitomagnetic tidal
field is taken to be stationary; in this case the tidal interaction
produces an internal velocity field that grows linearly with time. The
phenomenon was attributed to zero-frequency fluid modes, which do not 
provide a restoring force that would balance out the external tidal
forces and keep the fluid stationary. 

The analysis presented in Ref.~\cite{landry-poisson:15c} was performed
within a perturbative context in which the tidal forces are weak and
idealized as time-independent, and in which the body is only allowed
to rotate slowly. The calculations, however, were carried out in full
general relativity, and therefore incorporated all strong-field
effects in the interior of the compact body. The intrinsic complexity
of the computations prevented these authors from developing an
intuitive physical picture for the phenomenon, and kept them from
assessing its significance. In particular, the zero-frequency modes
were presented as a likely culprit for the phenomenon, but their 
precise role could not be ascertained.  

In this paper we provide the physics insights that were missing from
the original analysis. Our strategy is to approach the problem anew in
the framework of post-Newtonian gravity, assuming that 
the internal gravity of the compact body is not too strong. While the 
predictions of this analysis are likely to have limited accuracy from
a quantitative point of view, they will be qualitative robust, and 
they come with an intuitive understanding that was not easily
accessible in the general-relativistic treatment. We retain the 
assumption that the tidal forces are weak, but we no longer rely on
the stationary idealization; our tidal field can now vary with time,
on a time scale that's assumed to be long compared with the 
body's internal, hydrodynamical time scale. We also retain the
assumption that the body is rotating slowly.  

In this post-Newtonian setting, the generation of gravitomagnetic
tidal currents inside a rotating neutron star can be revealed with a 
relatively simple analysis. More importantly, the close analogy between
post-Newtonian gravity and Maxwell's electromagnetism provides us with
a strong intuitive basis. The phenomenon no longer looks so mysterious
when viewed in this particular way.   

We may now state the problem more precisely, and develop the intuition
behind the phenomenon. We consider a body of mass $M$, radius $R$, and
angular velocity $\bm{\Omega}$ immersed in a gravitomagnetic tidal
environment created by a remote companion of mass $M'$ moving with
velocity $v'$ on an orbit of radius $r'$. Just as an orbiting electric
charge would create a magnetic field, the orbiting companion creates a
gravitomagnetic field $\bm{B}$ around the body. The inhomogeneous
piece of this field, the one responsible for the tidal interaction,
scales as $GMv' \bm{x}/r^{\prime 3}$, in which $\bm{x}$ is the
position from the body's center-of-mass. The tidal field couples to
the body's rotational velocity $\bm{v} = \bm{\Omega} \times \bm{x}$
and creates, inside the body, a force density given by    
\begin{equation} 
\bm{f} =  \frac{1}{c^2} \rho \bm{v} \times \bm{B}, 
\end{equation} 
the gravitational analogue of the Lorentz force. Unlike the typical
situation encountered in Newtonian tides, or in general relativistic,  
gravitoelectric tides, this force is not balanced out by
pressure-gradient forces within the fluid. Instead, the
gravitomagnetic tidal forces act on the body and establish a velocity 
perturbation $\delta \bm{v}$ proportional to the time integral of the
tidal field. In the idealization of a time-independent $\bm{B}$, 
the velocity field would grow linearly in time, just as revealed in
Ref.~\cite{landry-poisson:15c}. In the more realistic case of a
time-changing $\bm{B}$, $\delta \bm{v}$ is modulated by the 
changes in the tidal environment. 

The root of the phenomenon is therefore an unbalanced force that
arises from the coupling between the gravitomagnetic tidal field and
the body's rotational velocity. These elementary considerations imply
that the velocity perturbation must scale as   
\begin{equation} 
\delta v = \frac{GM'v'\Omega R^2}{c^2 \omega' r^{\prime 3}}, 
\label{deltav_scale} 
\end{equation} 
where $\omega' := v'/r'$ is the orbital angular velocity, which
corresponds to the frequency of oscillation of the tidal field. In
this expression, the factor $GM'/r^{\prime 3}$ indicates 
that the effect is the result of a tidal interaction, the factor
$v'/c^2$ further reveals that it is a post-Newtonian, gravitomagnetic
effect, the factor $\Omega R$ shows that the effect results from the
coupling with the body's rotational velocity, the factor $R$ comes
from the scaling of $\bm{B}$ with the position relative to the 
body's center-of-mass, and the last factor $1/\omega'$ comes from the
time integral of the force density. 

The scale of the velocity perturbation can be re-expressed in terms of
the masses $M$ and $M'$, the body's rotation period $P := 2\pi/\Omega$
and radius $R$, and the orbital frequency $f' := \omega'/(2\pi)$. We
rely on Kepler's law $\omega^{\prime 2} = G(M+M')/r^{\prime 3}$ to
eliminate the orbital radius $r'$, and get 
\begin{equation} 
\delta v = \frac{(2\pi)^{7/3}  G^{1/3}}{c^2} \frac{M'}{(M+M')^{2/3}} 
\frac{R^2}{P} f^{4/3}.  
\end{equation} 
Inserting fiducial values for a typical binary system of neutron
stars near merger, we find that the scale of the velocity perturbation
is given by 
\begin{equation} 
\delta v = 2 \biggl( \frac{M'}{1.4\ M_\odot} \biggr) 
\biggl( \frac{2.8\ M_\odot}{M+M'} \biggr)^{2/3} 
\biggl( \frac{R}{12\ \mbox{km}} \biggr)^2 
\biggl( \frac{100\ \mbox{ms}}{P} \biggr) 
\biggl( \frac{f}{100\ \mbox{Hz}} \biggr)^{4/3}\ \mbox{km}/\mbox{s}. 
\end{equation} 
The amplitude of the tidal currents is measured in kilometers per
second, and should therefore be significant in these systems. 

In the remaining sections of the paper we give a precise statement of
these results and provide a complete derivation, taking the compact 
body to be a rigidly rotating perfect fluid, and considering generic tidal
environments. As we stated previously, our treatment is based on four
key assumptions. First, we take the gravitational field inside the
body to be sufficiently weak to permit a post-Newtonian approximation
carried out to the first order. Second, we assume that the tidal
perturbation is small and can be adequately described by a first-order
perturbative treatment. Third, we assume that the gravitomagnetic
tidal field changes on a time scale that is long compared with the
body's internal, hydrodynamical time scale. And fourth, we assume that
the body rotates slowly, so that all equations can be linearized with 
respect to the angular velocity $\Omega$.      

We begin in Sec.~\ref{sec:gm_tide} with a presentation of those
aspects of post-Newtonian gravity that are relevant for our
purposes. In particular, we introduce the vector potential $\bm{U}$
associated with the gravitomagnetic field, and describe a generic
gravitomagnetic tidal environment in terms of a symmetric-tracefree
tensor $\B_{ab}(t)$. In Sec.~\ref{sec:response} we introduce the
post-Newtonian version of Euler's equation, which governs the
behavior of a perfect fluid. We first integrate this equation for the
unperturbed configuration of a nonrotating star, and then switch on
the tidal field and the rotation. We find that the perturbation
equations take the same form as those governing a nonrotating 
fluid in Newtonian gravity, but with a driving force that 
couples the gravitomagnetic tidal field to the body's rotational
velocity. In Sec.~\ref{sec:Lagrangian} we recast the perturbation
equations in a convenient form involving a Lagrangian displacement
vector $\bm{\xi}$. This reformulation provides the basis for the
schematic mode analysis carried out in Sec.~\ref{sec:mode}, in which
we introduce the crucial zero-frequency modes and describe how they
can give rise to a velocity perturbation that behaves as in
Eq.~(\ref{deltav_scale}). 

In Sec.~\ref{sec:harmonics} we prepare the way for an actual
integration of the perturbation equations by expanding each variable
in spherical harmonics. Because the overall perturbation is a
composition of an $\ell = 1$ rotational perturbation with an 
$\ell =2$ tidal perturbation, the decomposition involves spherical
harmonics with $\ell = 1$, $\ell = 2$, and $\ell = 3$. These come in
two types, even-parity harmonics to represent scalars and vectors, and
odd-parity harmonics to represent pseudovectors. The decomposition
turns the perturbation equations into three decoupled sets of
equations, one set for each value of $\ell$. The explicit integration
of these equations is carried out in Sec.~\ref{sec:solution}. For
concreteness and simplicity we adopt a stellar model based on the
polytropic equation of state $p \propto \rho^2$, and the solution is
obtained with a mixture of analytical and numerical methods. The
precise expression of Eq.~(\ref{deltav_scale}) is provided by
Eq.~(\ref{wa_quadrupole}) and the following equations, as well as by 
Eq.~(\ref{wa_dipole}) and the following equations. The integration
reveals that the velocity field includes dipole ($\ell = 1$) and
quadrupole ($\ell=2$) components only; the expected octupole 
($\ell = 3$) contribution is absent because of a fortuitous
cancellation of the driving force at first post-Newtonian order. 

We return to the mode analysis in Sec.~\ref{sec:ZFmodes}, and convert
the schematic discussion of Sec.~\ref{sec:mode} into an actual method
to solve the perturbation equations for $\delta \bm{v}$. We confirm
that the solution constructed in Sec.~\ref{sec:solution} is indeed
generated by a degenerate family of zero-frequency modes, which can be
segregated into even-parity $g$-modes (those relevant for the
dipole piece of the velocity perturbation) and odd-parity $r$-modes
(which are relevant for the quadrupole piece of the
perturbation). We therefore validate the suggestion of
Ref.~\cite{landry-poisson:15c}, that zero-frequency 
modes are responsible for the gravitomagnetic tidal currents 
inside a rotating compact body.  

A number of additional results are worked out in appendices. In
Appendix \ref{app:Love} we calculate the post-Newtonian approximation
to the octupole, rotational-tidal Love number of a $p \propto \rho^2$
polytrope. In Appendix \ref{app:dipole} we justify the subtle boundary
conditions of the $\ell = 1$ perturbation equations at the stellar
surface. And finally, in Appendix \ref{app:CM} we show that the 
$\ell = 1$ acceleration field inside the body averages to a zero
overall acceleration for the body's center-of-mass.   

\section{Gravitomagnetic tidal field} 
\label{sec:gm_tide} 

Throughout this work we adopt the post-Newtonian approximation to 
relativistic gravity (see Chapter 8 of Poisson and Will's 
{\it Gravity} \cite{poisson-will:14}), based on the (Newtonian)
gravitoelectric potential $U$ and the gravitomagnetic vector potential
$U_a$. These satisfy the field equations 
\begin{equation} 
\nabla^2 U = -4\pi G \rho, \qquad 
\nabla^2 U_a = -4\pi G \rho v_a, 
\end{equation} 
in which $\rho$ is the mass density of the matter
distribution (denoted $\rho^*$ in {\it Gravity}), and $\bm{v}$ is its 
velocity field. The potentials are assumed to satisfy the harmonic
gauge condition 
\begin{equation} 
\partial_t U + \partial_a U^a = 0, 
\end{equation} 
and the matter variables satisfy the continuity equation 
\begin{equation} 
\partial_t \rho + \partial_a (\rho v^a) = 0. 
\end{equation} 
A complete description of post-Newtonian gravity also involves
an additional potential $\Psi$, which provides a correction of order
$(v/c)^2$ to the gravitoelectric potential. This potential, however,
is not required for our purposes in this work. 

We consider a rotating, self-gravitating body of mass $M$, radius
$R$, and angular velocity $\Omega$ immersed in a tidal environment
created by remote objects. Our considerations are limited to a
spherical domain $\M$ described by $0 < r < r_{\rm max}$, where $r$ is
the distance to the body's center-of-mass, and $r_{\rm max} > R$ is a
maximum distance from the body. This domain includes the body, but it
excludes the remote objects that create the tidal environment.  

A Newtonian description of the tidal environment is provided in
Sec.~2.5 of {\it Gravity}. The gravitoelectric potential is decomposed
into a body piece $U^{\rm body}$ and an external piece $U^{\rm ext}$,
and since the sources of the external potential are outside $\M$, it
must satisfy Laplace's equation $\nabla^2 U^{\rm ext} = 0$. Assuming
that the scale of variation of the external potential is large
compared with $R$, we express it as the Taylor expansion 
$U^{\rm ext}(x^a) = U^{\rm ext}(0) 
+ g_{a} x^a - \frac{1}{2} \E_{ab} x^a x^b + \cdots$, where 
$g_a := \partial_a U^{\rm ext}(0)$ and 
$\E_{ab} := -\partial_{ab} U^{\rm ext}(0)$, with $x^a$ denoting the
position relative to the body's center-of-mass. The leading term is an
irrelevant constant, the linear term is responsible for the motion of
the center-of-mass, and the additional terms are responsible
for the tides. To leading order in the tidal interaction, we have that 
\begin{equation} 
U^{\rm tidal} = -\frac{1}{2} \E_{ab}\, x^a x^b, 
\end{equation} 
with the gravitoelectric tidal quadrupole moment $\E_{ab}(t)$
providing a complete characterization of the tidal environment. The
definition of $\E_{ab}$ implies that this tensor is symmetric, and the
field equation $\nabla^2 U^{\rm tidal} = 0$ further implies that it
is tracefree: $\E_{aa} = 0$. The tidal moment is therefore a
symmetric-tracefree (STF) tensor that possesses 5 independent
components.  

We shall assume that the time scale of variation of $\E_{ab}$ is very
long compared with the internal, hydrodynamical time scale of the
body, which is comparable to $\sqrt{R^3/GM}$. This will allow us to
neglect the time derivatives of the tidal gravitoelectric potential.  

The gravitomagnetic potential $U_a$ also contributes to the tidal
environment. We examine the external piece of this potential, which is
sourced by the remote objects; it satisfies 
$\nabla^2 U^{\rm ext}_a = 0$ in addition to the gauge condition 
$\partial_a U^{\rm ext}_a = 0$, in which we have neglected the term  
$\partial_t U^{\rm ext}$, as motivated previously. A gauge
transformation $U^{\rm ext}_a \to U^{\rm ext}_a + \partial_a f$
preserves the gauge condition provided that $f$ satisfies
Laplace's equation. 

We perform a Taylor expansion of the external gravitomagnetic
potential, and discard the irrelevant constant term and the
linear term responsible for the center-of-mass motion. The tidal
potential therefore leads with $U^{\rm tidal}_a = A_{abc} x^b x^c$,
with $A_{abc}(t)$ defined to be symmetric in the last two indices;
this tensor contains 18 independent components. The gauge condition
gives rise to the three constraints $A_{aab} = 0$, and the number of
independent components reduces to 15. An additional reduction is made
possible by a gauge transformation generated by $f = C_{abc} x^a x^b
x^c$, where $C_{abc}$ is completely symmetric by virtue of its
definition, and satisfies $C_{aab} = 0$ by virtue of the requirement
that $\nabla^2 f = 0$. There are 7 independent components in
$C_{abc}$, and these can be chosen to eliminate an equal number of
components in $A_{abc}$; the count is therefore reduced to 8
independent components. Finally, the field equations 
$\nabla^2 U^{\rm tidal}_a = 0$ introduce 3 new constraints, and the
number of independent components has finally settled to 5. These can
be encoded in the STF tensor $\B_{ab}(t)$, and it can be verified that  
\begin{equation} 
U^{\rm tidal}_a = -\frac{1}{6} \epsilon_{abc} \B^c_{\ d}\, x^b x^d
\label{Ua_tidal} 
\end{equation} 
satisfies the gauge condition (because $\B_{ab}$ is symmetric) and the 
field equations (because $\B_{ab}$ is tracefree). Equation
(\ref{Ua_tidal}), therefore, provides a correct description of a
gravitomagnetic tidal potential \cite{zhang:86}.  

The gravitomagnetic tidal quadrupole moment $\B_{ab}$ can be
expressed as $\B_{ab} = 2 \epsilon_{cd(a} \partial_{b) c} 
U^{\rm ext}_d(0)$, in terms of second derivatives of the external
potential evaluated at $x^a = 0$. For a tidal environment created by a
single companion of mass $M'$ moving with velocity $\bm{v'}$ at a
position $\bm{r'}$ from the body, $U^{\rm ext}_a = G M' v'_a/r'$, and  
\begin{equation} 
\B_{ab} = \frac{6 G M'}{r^{\prime 3}} 
( \bm{n'} \times \bm{v'} )_{(a} n'_{b)}, 
\label{Bab_explicit} 
\end{equation} 
where $\bm{n'} := \bm{r'}/r'$. When the companion moves on a circular
orbit of radius $r'$ in the $x$-$y$ plane of the coordinate system,
the nonvanishing components of $\B_{ab}$ are 
\begin{equation} 
B_{xz} = \frac{3GM' v'}{r^{\prime 3}} \cos\Phi, \qquad 
B_{yz} = \frac{3GM' v'}{r^{\prime 3}} \sin\Phi,
\label{Bab_circular} 
\end{equation} 
where $\Phi := \omega' t$ is the orbital phase, with  
\begin{equation} 
\omega' := \sqrt{ \frac{G(M+M')}{r^{\prime 3}} }
\label{orbital_omega} 
\end{equation} 
denoting the orbital angular velocity, related to the orbital velocity
by $v' = r' \omega'$.  

\section{Body's response to a gravitomagnetic tidal field} 
\label{sec:response} 

In this section we derive the equations that govern the response of
a rotating body to the tidal gravitomagnetic potential of
Eq.~(\ref{Ua_tidal}). We ignore the influence of the gravitoelectric
tidal field, which gives rise to the well-understood Newtonian tides
(see, for example, Sec.\ 2.5 of {\it Gravity}
\cite{poisson-will:14}). The body is modelled 
as a perfect fluid with a zero-temperature equation of state of the
form $p = p(\rho)$, with $p$ denoting the pressure. Its response is
determined by the post-Newtonian version of Euler's equation,
displayed in Eq.~(8.119) of {\it Gravity},  
\begin{align} 
\rho \frac{d v_a}{dt} &= -\partial_a p + \rho \partial_a U 
+ \frac{1}{c^2} \biggl[ \biggl( \frac{1}{2} v^2 + U + \Pi 
+ \frac{p}{\rho} \biggr) \partial_a p - v_a \partial_t p \biggr] 
\nonumber \\ & \quad \mbox{} 
+ \frac{\rho}{c^2} \Bigl[ (v^2 - 4U) \partial_a U 
- v_a \bigl( 3\partial_t U + 4 v^b \partial_b U \bigr) 
+ 4 \partial_t U_a + 4 v^b \bigl( \partial_b U_a 
- \partial_a U_b \bigr)  + \partial_a \Psi \Bigr] 
+ O(c^{-4}), 
\label{euler} 
\end{align} 
where $d/dt := \partial_t + v^b \partial_b$ is the convective time
derivative, $\Pi$ is the fluid's internal energy per unit mass, and
$\Psi$ is the post-Newtonian potential mentioned previously. 

We begin with a nonrotating body in an unperturbed state, in the
absence of a perturbing tidal field. In this context the body is
static and spherically symmetric, and its structure is determined by
the equations $\partial_a p = \rho \partial_a U + O(c^{-2})$ and 
$\nabla^2 U = -4\pi G \rho$. We allow ourselves to neglect all
post-Newtonian corrections to the structure equations, which take, in
this approximation, the explicit form 
\begin{equation} 
\frac{dp}{dr} = \rho \frac{dU}{dr} = -\frac{G m \rho}{r^2}, \qquad 
\frac{dm}{dr} = 4\pi r^2 \rho,  
\label{hydro_eq} 
\end{equation} 
with $m(r)$ denoting the internal mass function. 

We next switch on the gravitomagnetic tidal field, but keep the 
body nonrotating. We assume that the tidal field is small and creates
a change in the fluid configuration that can adequately be described
by a first-order perturbative treatment. We further assume that
$\B_{ab}(t)$ changes on a time scale that is long compared with the
time scale of internal hydrodynamical processes. And we assume that
$\B_{ab}(t \to -\infty) \to 0$, so that the body's initial state is
the unperturbed state described previously. As we shall see presently,
the fluid acquires a velocity field $v_a$ as a result of the tidal
interaction, and the total gravitomagnetic potential is 
$U_a = U_a^{\rm body} + U_a^{\rm tidal}$, with the body piece
satisfying $\nabla^2 U_a^{\rm body} = -4\pi G \rho v_a$. 

The gravitomagnetic tidal perturbation keeps all scalar quantities
(such as $\rho$, $p$, $U$, $\Pi$, and $\Psi$) unchanged to first order
in perturbation theory. The reason is tied to their behavior under a
parity transformation, in which $x^a \to -x^a$. Scalar quantities are
not affected by the transformation, while a vector such as $U_a$
changes sign. Now, Eq.~(\ref{Ua_tidal}) reveals that 
$\B_{ab} \to -\B_{ab}$ under the transformation 
($\epsilon_{abc}$ is unaffected), and the gravitomagnetic tidal moment 
therefore behaves as a pseudotensor. Because a perturbation in a
scalar quantity would have to be proportional to $\B_{ab} x^a x^b$ to
be a scalar, and because this does change sign under a parity
transformation (it is a pseudoscalar instead of a true scalar), we
must rule out such perturbations.    

With $\delta \rho = \delta p = \delta U = 0$, the post-Newtonian
Euler equation implies that $\rho dv_a/dt = O(c^{-2})$, so that the
velocity field must be of order $c^{-2}$. This immediately implies
that $U^{\rm body}_a = O(c^{-2})$, so that
\begin{equation} 
U_a = U_a^{\rm tidal} + O(c^{-2}) 
= -\frac{1}{6} \epsilon_{abp} \B^p_{\ c}\, x^b x^c 
+ O(c^{-2}). 
\end{equation} 
These observations give rise to a huge simplification in
Eq.~(\ref{euler}). A careful examination of the equation, neglecting
all terms that are beyond first order in the perturbation, and
all terms that are beyond the first post-Newtonian order, reveals that
it reduces to $\partial_t (v_a - 4 U_a/c^2) = O(c^{-4})$. Because  
the fluid is assumed to be unperturbed initially, the time
independence of $v_a - 4 U_a/c^2$ guarantees that 
\begin{equation} 
v_a = \frac{4}{c^2} U_a + O(c^{-4})
\label{irrot} 
\end{equation} 
at all times. The gravitomagnetic tidal interaction therefore creates
a velocity field within the fluid, which gradually builds up as the
tidal field is switched on. This velocity field is required by the
relativistic circulation theorem \cite{shapiro:96, favata:06}.  

We now allow the body to rotate. For simplicity we take the body
to rotate rigidly with an angular velocity $\Omega$. Aligning the
rotation axis with the $z$-direction, we define the vector 
$\Omega^a = [0,0,\Omega]$ and the rotational velocity field is 
\begin{equation} 
v_a^{\rm rot} = \epsilon_{abc} \Omega^b x^c. 
\end{equation} 
We assume that $\Omega$ is sufficiently small that centrifugal effects
on the body's structure can be neglected. This amounts to demanding
that $\Omega^2 \ll G m(r)/r^3$ throughout the body, and the assumption
allows us to work to first order in $v_a^{\rm rot}$. The rotating body
is perturbed by the gravitomagnetic tidal field, and the coupling
between $v_a^{\rm rot}$ and $U_a^{\rm tidal}$ ensures that in addition
to the fluid's velocity field, $\rho$, $p$, and $U$ also acquire
perturbations; parity considerations no longer rule them out, because
the pseudovector $v_a^{\rm rot}$ can combine with the pseudotensor 
$\B_{ab}$ and the vector $x^a$ to form scalar quantities. Because the
interaction with the gravitomagnetic tidal field is a post-Newtonian
effect, all perturbations will be post-Newtonian quantities of order
$c^{-2}$.   

We let $\rho \to \rho + \delta \rho$, $p \to p + \delta p$, 
$U \to U + \delta U$, $v_a \to v_a^{\rm rot} + \delta v_a$, and 
$U_a = U_a^{\rm tidal}$ in the post-Newtonian Euler equation, and
expand the equation to first order in all perturbations, taking into
account the important fact that these are all of order $c^{-2}$. After
simplification we arrive at  
\begin{equation} 
\partial_t \delta v_a + v^b \partial_b \delta v_a 
+ (\partial_b v_a) \delta v^b - P_a = \frac{4}{c^2} W_a + O(c^{-4}), 
\label{Eulerpert1} 
\end{equation} 
where 
\begin{equation} 
P_a := -\frac{1}{\rho} \partial_a \delta p 
+ \frac{\delta \rho}{\rho} \partial_a U 
+ \partial_a \delta U 
\label{P_def} 
\end{equation} 
and 
\begin{equation} 
W_a := \partial_t U_a + v^b \bigl( \partial_b U_a 
- \partial_a U_b \bigr). 
\end{equation} 
To simplify the notation we let $v_a \equiv v_a^{\rm rot}$ and $U_a
\equiv U_a^{\rm tidal}$, as given by Eq.~(\ref{Ua_tidal}). 

Equation (\ref{Eulerpert1}) is the starting point of the perturbative
analysis. Because $\delta v_a = 4U_a/c^2$ when $\Omega = 0$, we write 
\begin{equation} 
\delta v_a = \frac{4}{c^2} U_a + w_a 
\label{w_def} 
\end{equation} 
and consider $w_a$ to be a post-Newtonian quantity of order
$\Omega$. Making the substitution in Eq.~(\ref{Eulerpert1}) and
neglecting all terms beyond first order in $\Omega$, we arrive at the
simpler equation 
\begin{equation} 
\partial_t w_a - P_a = -\frac{4}{c^2} A_a + O(c^{-4}), 
\label{Eulerpert2} 
\end{equation} 
where 
\begin{equation} 
A_a := ( \partial_b v_a ) U^b + v^b \partial_a U_b. 
\label{A_def} 
\end{equation} 
The left-hand side of Eq.~(\ref{Eulerpert2}) features the familiar 
linearization of the Newtonian Euler equation for the perturbation of
a nonrotating fluid, and the right-hand side features a post-Newtonian
driving force $A_a$ that originates from the coupling between the
gravitomagnetic tidal field $U_a$ and the rotational velocity $v_a$.  
The equation is mathematically equivalent to one describing a
nonrotating body perturbed by a prescribed driving force. 

Equation (\ref{Eulerpert2}) must be supplemented by Poisson's equation  
\begin{equation} 
\nabla^2 \delta U = -4\pi G \delta \rho 
\label{Poisson} 
\end{equation} 
for the perturbation of the Newtonian potential, the continuity
equation expressing mass conservation, and an equation of state for
the perturbed fluid. Throughout this work we shall assume that the
perturbed fluid satisfies the same equation of state as the
unperturbed fluid. 

\section{Lagrangian description of the fluid perturbation} 
\label{sec:Lagrangian} 

The perturbative treatment of the previous section was couched in the
language of Eulerian perturbations, with a perturbation such as 
$\delta \rho$ comparing the perturbed and unperturbed fluids at the
same spatial position. We next introduce a Lagrangian description, in
which a perturbation such as $\Delta \rho$ compares the perturbed and
unperturbed fluids at the same fluid element. The relation between the
two descriptions is provided by the Lagrangian displacement vector
$\xi^a$, which gives the position of a given fluid element in the
perturbed fluid relative to its position in the unperturbed fluid. The
Eulerian and Lagrangian perturbations are related by 
$\Delta = \delta + \xi^a \partial_a$. In the Lagrangian description,
the perturbation in the velocity field is $\Delta v^a = d\xi^a/dt$,
mass conservation is embodied in $\Delta \rho = -\rho \partial_a
\xi^a$, and with the assumption placed earlier on the equation of
state, $\Delta p = (dp/d\rho)\, \Delta \rho$. In terms of
Eulerian variations, we have  
\begin{equation} 
\delta v^a = \partial_t \xi^a + v^b \partial_b \xi^a 
- \xi^b \partial_b v^a, \qquad 
\delta \rho = -\partial_a(\rho \xi^a),
\label{euler_vs_xi} 
\end{equation} 
and $\delta p = (dp/d\rho)\, \delta \rho$. 

We have seen that $\delta v_a = 4U_a/c^2$ when the body is nonrotating
($v_a = 0$). In this case $\delta v_a = \partial_t \xi^a$, and to
reflect the change of variables of Eq.~(\ref{w_def}) to account for
the rotation, we write 
\begin{equation} 
\xi^a(t,x^b) = \frac{4}{c^2} \int^t U^a(t',x^b)\, dt' + \zeta^a(t, x^b), 
\label{zeta_def} 
\end{equation} 
where $\zeta^a$ is a post-Newtonian quantity of order $\Omega$. Making
the substitution in Eq.~(\ref{euler_vs_xi}) yields 
\begin{equation} 
w_a (t,x^b) = \partial_t \zeta_a (t,x^b) 
+ \frac{4}{c^2} \int^t C_a (t',x^b)\, dt'  
 \label{euler_vs_zeta1} 
\end{equation} 
with  
\begin{equation} 
C_a := v^b \partial_b U_a - U^b \partial_b v_a, 
\label{C_def}   
\end{equation} 
and 
\begin{equation} 
\delta \rho = -\partial_a(\rho \zeta^a). 
\label{euler_vs_zeta2} 
\end{equation} 
There is no integral term in the last equation, because 
$\partial_a U^a = 0$ and $U^a \partial_a \rho = r^{-1} (d\rho/dr) 
x^a U_a  = 0$. The Euler equation (\ref{Eulerpert2}) becomes 
\begin{equation} 
\partial_{tt} \zeta_a - P_a = -\frac{4}{c^2} B_a + O(c^{-4}), 
\label{Eulerpert3} 
\end{equation} 
with 
\begin{equation} 
B_a := v^b \bigl( \partial_a U_b + \partial_b U_a \bigr). 
\label{B_def} 
\end{equation} 
The equation is again supplemented by
Eq.~(\ref{Poisson}). The continuity equation has already been 
incorporated in Eq.~(\ref{euler_vs_zeta2}), and as we have seen, the
pressure perturbation is given by $\delta p = (dp/d\rho)\, \delta \rho$. 

The integral term in Eq.~(\ref{euler_vs_zeta1}) suggests that $w_a$
might be expected to grow in time, on a short time scale compared with
the scale of variation of the tidal potential. We shall see this
expectation confirmed when we construct the solution to the
perturbation equations.     

\section{Mode analysis} 
\label{sec:mode} 

Equation (\ref{Eulerpert3}) can be integrated by performing a mode
analysis. We examine the homogeneous equation, $\partial_{tt} \zeta_a
- P_a = 0$ with $P_a$ given by Eq.~(\ref{P_def}), and recognize that
by virtue of Eqs.~(\ref{Poisson}) and (\ref{euler_vs_zeta2}), $P_a$ is
a linear functional of $\zeta_a$. We express it as 
$P_a  = -\LL_a^{\ b}\, \zeta_b$, in which $\LL_a^{\ b}$ is an
integro-differential operator that is known to be
self-adjoint with respect to the measure $\rho\, d^3x$ 
\cite{chandrasekhar:64}. Writing 
\begin{equation} 
\zeta_a(t,x^b) = f_a(x^b) e^{-i\omega t}, 
\end{equation} 
we find that the homogeneous equation
turns into the eigenvalue equation $\LL_a^{\ b} f_b = \omega^2 f_a$
for the modes $f_a$. With $\LL_a^{\ b}$ self-adjoint, the eigenvalues
$\omega^2$ are guaranteed to be real, and modes with different
frequencies are guaranteed to be orthogonal. Introducing the mode
label $\lambda$, we denote the eigenvalues $\omega_\lambda$, the
corresponding mode functions $f^a_\lambda$, and the orthogonality
property is expressed by
\begin{equation} 
\int \rho\, \bm{f}_\lambda \cdot \bm{f}_{\lambda'}\, d^3x 
= N_\lambda \delta_{\lambda\lambda'}, 
\end{equation} 
with $N_\lambda$ denoting the normalization of each mode. The
spectrum of $\LL_a^{\ b}$ is also known to include an infinitely
degenerate set of zero-frequency modes that satisfy 
$\LL_a^{\ b} f_b = 0$. These are necessarily orthogonal to those 
with nonzero frequency, and they can be made mutually
orthogonal by implementing a Gram-Schmidt procedure. We label the 
zero-frequency modes with the index $I$, and express their
orthogonality as  
\begin{equation} 
\int \rho\, \bm{f}_I \cdot \bm{f}_{I'}\, d^3x 
= N_I \delta_{II'}.  
\end{equation} 
We take it for granted that the entire collection of modes
$f_\lambda^a$ and $f_I^a$ forms a complete set. Completeness under
certain assumptions was proved by Beyer and Schmidt
\cite{beyer-schmidt:95}.  

Returning to Eq.~(\ref{Eulerpert3}), we decompose $\zeta_a$ and $B_a$
into modes, 
\begin{subequations}
\label{mode_expansion} 
\begin{align} 
\zeta^a(t,x^b) &= \sum_I z_I(t) f_I^a(x^b) 
+ \sum_\lambda z_\lambda(t) f_\lambda^a(x^b),  \\  
B^a(t,x^b) &= \sum_I B_I(t) f_I^a(x^b) 
+ \sum_\lambda B_\lambda(t) f_\lambda^a(x^b), 
\end{align} 
\end{subequations}   
with mode amplitudes given by 
\begin{equation} 
z_I = \frac{1}{N_I} \int \rho\, \bm{\zeta} \cdot
\bm{f}_I\, d^3x, \qquad 
B_I = \frac{1}{N_I} \int \rho\, \bm{B} \cdot
\bm{f}_I\, d^3x  
\label{mode_amplitudes} 
\end{equation} 
and analogous equations for $z_\lambda$ and $B_\lambda$. We make
the substitutions, invoke the mode equation and the orthogonality
relations, and obtain   
\begin{equation} 
\ddot{z}_I = -\frac{4}{c^2} B_I, \qquad 
\ddot{z}_\lambda + \omega^2_\lambda z_\lambda 
= -\frac{4}{c^2} B_\lambda, 
\end{equation} 
with an overdot indicating differentiation with respect to $t$. Each
mode is seen to behave as a driven harmonic oscillator, and with the
assumption that the fluid begins in an unperturbed state at 
$t =-\infty$, the solutions are  
\begin{equation} 
z_I(t) = -\frac{4}{c^2}\int_{-\infty}^t (t-t') B_I(t')\, dt', \qquad 
z_\lambda(t) = -\frac{4}{\omega_\lambda c^2} \int_{-\infty}^t 
dt'\, B_\lambda(t') \sin \bigl[ \omega_\lambda(t-t') \bigr]. 
\end{equation} 
While the modes $f^a_\lambda$ give rise to oscillating contributions
to $\zeta_a$, the zero-frequency modes $f^a_I$ produce a growing
contribution that can be expressed as 
\begin{equation} 
\zeta_a^{\rm grow}(t,x^b) = -\frac{4}{c^2} 
\int_{-\infty}^t (t-t') B_a^{\rm zf}(t',x^b)\, dt', 
\label{zeta_grow} 
\end{equation} 
where $B_a^{\rm zf}(t,x^b) := \sum_I B_I(t) f_a^I(x^b)$ is the
projection of $B_a$, as defined by Eq.~(\ref{B_def}), onto the
zero-frequency subspace. This gives rise to a second growing
contribution to the velocity field, in addition to the one already
displayed in Eq.~(\ref{euler_vs_zeta1}). The growing piece of the
velocity field is then given by 
\begin{equation} 
w_a^{\rm grow}(t,x^b) = \frac{4}{c^2} \int^t 
\bigl[ C_a(t',x^b) - B_a^{\rm zf}(t',x^b) \bigr]\, dt', 
\label{w_grow} 
\end{equation} 
with $C_a$ defined by Eq.~(\ref{C_def}). The zero-frequency modes  
$f^a_I$ play a crucial role in the response of a fluid body to a
gravitomagnetic tidal field, giving rise to a velocity perturbation
that can be expected to grow in time. We shall examine them in  
detail in Sec.~\ref{sec:ZFmodes}.   

\section{Spherical-harmonic decomposition} 
\label{sec:harmonics} 

The mode analysis carried out in Sec.~\ref{sec:mode} supplies us with
a powerful conceptual framework to analyze the perturbation equation 
(\ref{Eulerpert3}), and it provides us with an expectation that thanks
to the zero-frequency modes, the solution $\zeta_a$ will contain
growing terms. To make further progress we return to
Eq.~(\ref{Eulerpert3}) and perform a decomposition in spherical
harmonics. 

\begin{table}
\caption{Spherical-harmonic functions $Y^{\ell\m}$. The functions are
  real, and they are listed for the relevant modes $l=1$ (dipole),
  $l=2$ (quadrupole), and $l=3$ (octupole). The abstract index $\m$
  describes the dependence of these functions on the angle $\phi$; for
  example $Y^{\ell,2s}$ is proportional to $\sin2\phi$.} 
\begin{ruledtabular} 
\begin{tabular}{l} 
$ Y^{1,0} = \cos\theta $ \\ 
$ Y^{1,1c} = \sin\theta \cos\phi $ \\ 
$ Y^{1,1s} = \sin\theta \sin\phi $ \\ 
\\ 
$ Y^{2,0} = 1-3\cos^2\theta $ \\ 
$ Y^{2,1c} = 2\sin\theta\cos\theta\cos\phi $ \\ 
$ Y^{2,1s} = 2\sin\theta\cos\theta\sin\phi $ \\ 
$ Y^{2,2c} = \sin^2\theta\cos 2\phi $ \\ 
$ Y^{2,2s} = \sin^2\theta\sin 2\phi $ \\ 
\\
$ Y^{3,0} = \cos\theta(3-5\cos^2\theta) $ \\  
$ Y^{3,1c} = \frac{3}{2} \sin\theta(1-5\cos^2\theta)\cos\phi $ \\  
$ Y^{3,1s} = \frac{3}{2} \sin\theta(1-5\cos^2\theta)\sin\phi $ \\  
$ Y^{3,2c} = 3\sin^2\theta\cos\theta \cos 2\phi $ \\  
$ Y^{3,2s} = 3\sin^2\theta\cos\theta \sin 2\phi $ \\  
$ Y^{3,3c} = \sin^3\theta\cos 3\phi $ \\  
$ Y^{3,3s} = \sin^3\theta\sin 3\phi $
\end{tabular}
\end{ruledtabular} 
\label{tab:Ylm} 
\end{table} 

\begin{table}
\caption{Spherical-harmonic coefficients of tidal potentials.}   
\begin{ruledtabular} 
\begin{tabular}{l} 
$ \Bq_0 = \frac{1}{2} (\B_{11} + \B_{22}) $ \\ 
$ \Bq_{1c} = \B_{13} $ \\ 
$ \Bq_{1s} = \B_{23} $ \\ 
$ \Bq_{2c} = \frac{1}{2} ( \B_{11} - \B_{22}) $ \\ 
$ \Bq_{2s} = \B_{12} $ \\ 
\\ 
$ \Kd_0 = \K_3 = -2\Omega \Bq_0 $ \\ 
$ \Kd_{1c} = \K_1 = \Omega \Bq_{1c} $ \\ 
$ \Kd_{1s} = \K_2 = \Omega \Bq_{1s} $ \\ 
\\ 
$ \Ko_0 = \frac{1}{2} (\K_{113} + \K_{223}) 
= \frac{3}{5} \Omega \Bq_{0} $ \\ 
$ \Ko_{1c} = \frac{1}{2} (\K_{111} + \K_{122}) 
= -\frac{4}{15} \Omega \Bq_{1c} $ \\  
$ \Ko_{1s} = \frac{1}{2} (\K_{112} + \K_{222}) 
= -\frac{4}{15} \Omega \Bq_{1s} $ \\  
$ \Ko_{2c} = \frac{1}{2} (\K_{113} - \K_{223}) 
= \frac{1}{3} \Omega \Bq_{2c} $ \\  
$ \Ko_{2s} = \K_{123} = \frac{1}{3} \Omega \Bq_{2s} $ \\  
$ \Ko_{3c} = \frac{1}{4} (\K_{111} - 3\K_{122}) = 0 $ \\ 
$ \Ko_{2s} = \frac{1}{4} (3\K_{112} - \K_{222}) = 0 $ \\
\\
$ \Bhatq_0 = \frac{1}{2} (\hat{\B}_{11} + \hat{\B}_{22}) = 0$ \\ 
$ \Bhatq_{1c} = \hat{\B}_{13} = -\Omega \Bq_{1s} $ \\ 
$ \Bhatq_{1s} = \hat{\B}_{23} = \Omega \Bq_{1c} $ \\ 
$ \Bhatq_{2c} = \frac{1}{2} ( \hat{\B}_{11} - \hat{\B}_{22}) 
= -2\Omega \Bq_{2s} $ \\ 
$ \Bhatq_{2s} = \hat{\B}_{12} = 2 \Omega \Bq_{2c} $ 
\end{tabular}
\end{ruledtabular} 
\label{tab:coeffs} 
\end{table} 
 
To prepare the way for this decomposition we rely on Sec.~II of
Ref.~\cite{poisson:15} --- see also Sec.~II of Ref.~\cite{landry-poisson:15a}
--- and construct tidal potentials that form an irreducible
basis in which to decompose the driving force $B_a$ displayed in
Eq.~(\ref{B_def}). We first introduce the spherical coordinates
$(r,\theta^A)$, with $\theta^A = (\theta,\phi)$, which are related to
the Cartesian coordinates by $x^a = r n^a$, with $n^a :=
[\sin\theta\cos\phi, \sin\theta\sin\phi, \cos\theta]$. We also
introduce the spherical-harmonic functions $Y^{\ell \m}(\theta^A)$
displayed in Table~\ref{tab:Ylm}; these are defined to be real
functions, they are not normalized in the usual way, and the label
$\m$ describes their dependence on $\phi$. The association  
\begin{equation} 
\B_{ab} n^a n^b = \sum_\m \Bq_\m Y^{2,\m}
\end{equation} 
allows us to package the five independent components of
$\B_{ab}$ into the five harmonic coefficients $\Bq_\m$. The
superscript $\sf q$ stands for ``quadrupole'', and the explicit
relations between $\B_{ab}$ and $\Bq_\m$ are listed in
Table~\ref{tab:coeffs}.  

The pseudovector $\Omega_a$ and pseudotensor $\B_{ab}$ can be combined 
to form the vector and symmetric-tracefree (STF) tensor  
\begin{equation} 
\K_a := \B_{ab} \Omega^b, \qquad 
\K_{abc} := \B_{\langle a b} \Omega_{c\rangle}, 
\label{K_def} 
\end{equation} 
in which the angular brackets instruct us to symmetrize all indices
and remove all traces. They can also be combined into the STF
pseudotensor  
\begin{equation} 
\hat{\B}_{ab} := 2 \Omega^c \epsilon_{cd(a} \B^d_{\ b)}. 
\label{Bhat_def} 
\end{equation}  
The associations 
\begin{equation} 
\K_a n^a = \sum_\m \Kd_\m Y^{1,\m}, \qquad 
\K_{abc} n^a n^b n^c = \sum_\m \Ko_\m Y^{3,\m}, \qquad  
\hat{\B}_{ab} n^a n^b = \sum_\m \Bhatq_\m Y^{2,\m}
\label{assocs} 
\end{equation} 
define the harmonic coefficients $\Kd_\m$, $\Ko_\m$, and $\Bhatq_\m$, 
which are given explicitly in Table~\ref{tab:coeffs}. We may note that 
\begin{equation} 
\Kd_\m = \fd_\m\, \Omega \Bq_\m, \qquad 
\Ko_\m = \fo_\m\, \Omega \Bq_\m, 
\label{Kd_Ko} 
\end{equation} 
where the numbers $\fd_\m$ and $\fo_\m$ can be extracted from the
table. The superscripts $\sf d$ and $\sf o$ stand for ``dipole'' and
``octupole'', respectively. 

The tidal potentials are divided into scalar and vector
potentials. For our purposes here, a ``scalar'' is a quantity that
stays invariant under a transformation of the angular coordinates
$\theta^A$, while a ``vector'' is a quantity that transforms as a
one-form under this transformation. The scalar potentials can be
decomposed into scalar harmonics $Y^{\ell\m}$, but the vector
potentials require the even-parity vector harmonics 
\begin{equation} 
Y_A^{\ell\m} := \partial_A Y^{\ell\m}
\label{even_harm} 
\end{equation} 
and the odd-parity vector harmonics 
\begin{equation} 
X_A^{\ell\m} := -\epsilon_A^{\ B} \partial_B Y^{\ell\m},  
\label{odd_harm}  
\end{equation}  
in which $\epsilon_A^{\ B}$ is the Levi-Civita tensor on the unit 
2-sphere, with nonvanishing components $\epsilon_\theta^{\ \phi} 
= 1/\sin\theta$ and $\epsilon_\phi^{\ \theta} = -\sin\theta$. 

We may now state the definition of the tidal potentials. They are
given by 
\begin{subequations} 
\begin{align} 
&\Kd := \sum_\m \Kd_\m\, Y^{1,\m}, \qquad 
\Kd_A := \sum_\m \Kd_\m\, Y^{1,\m}_A, 
\label{Kd_pot}  \\
&\Ko := \sum_\m \Ko_\m\, Y^{3,\m}, \qquad 
\Ko_A := \frac{1}{3} \sum_\m \Ko_\m\, Y^{3,\m}_A,  
\label{Ko_pot} \\ 
&\Bhatq_A := \frac{1}{2} \sum_\m \Bhatq_\m\, X^{2,\m}_A
= -\frac{1}{2} \Omega \sum_\m \Bq_\m\, \partial_\phi X^{2,\m}_A, 
\label{Bhatq_pot} 
\end{align} 
\end{subequations} 
and they can be used as a basis to decompose the driving force $B_a$
displayed in Eq.~(\ref{B_def}). Simple manipulations reveal that 
\begin{subequations} 
\label{B_decomposed} 
\begin{align} 
& B_r := B_a n^a = \frac{1}{10} r^2 \Kd - \frac{1}{6} r^2 \Ko, 
\\ 
& B_A := r B_a \partial_A n^a = \frac{1}{5} r^3 \Kd_A 
- \frac{1}{6} r^3 \Ko_A + \frac{1}{9} r^3 \Bhatq_A. 
\end{align} 
\end{subequations} 
The vector $C_a$ defined by Eq.~(\ref{C_def}) can also be decomposed
in this basis. Here we find that 
\begin{equation} 
C_r := C_a n^a = 0, \qquad 
C_A := r C_a \partial_A n^a = \frac{1}{6} r^3 \Bhatq_A. 
\label{C_decomposed} 
\end{equation}

The fact that $B_r$ can be decomposed in spherical harmonics with 
$\ell = 1$ and $\ell = 3$ implies that all scalar perturbations can be
decomposed in a similar way. And the fact that $B_A$ can be decomposed
in even-parity vector harmonics with  $\ell = (1, 3)$ and in odd-parity
harmonics with $\ell = 2$ ensures that all vector perturbations can be
decomposed in the same way. We therefore write  
\begin{subequations} 
\begin{align} 
\delta p &= \Omega \sum_\m \fd_\m\, \pd(t,r)\, Y^{1,\m} 
+ \Omega \sum_\m \fo_\m\, \po(t,r)\, Y^{3,\m},
\label{deltap_spharm} \\  
\delta U &= \Omega \sum_\m \fd_\m\, \Ud(t,r)\, Y^{1,\m} 
+ \Omega \sum_\m \fo_\m\, \Uo(t,r)\, Y^{3,\m}, 
\label{deltaU_spharm} \\  
\zeta_r &= \Omega \sum_\m \fd_\m\, \zetad_r(t,r)\, Y^{1,\m}  
+ \Omega \sum_\m \fo_\m\, \zetao_r(t,r)\, Y^{3,\m}, 
\label{zetar_spharm} \\  
w_r &= \Omega \sum_\m \fd_\m\, \wwd_r(t,r)\, Y^{1,\m}  
+ \Omega \sum_\m \fo_\m\, \wwo_r(t,r)\, Y^{3,\m} 
\label{wr_spharm} \\  
\end{align} 
\end{subequations} 
as well as 
\begin{subequations} 
\begin{align} 
\zeta_A &= \Omega \sum_\m \fd_\m\, \zetad(t,r)\, Y^{1,\m}_A 
+ \frac{1}{3} \Omega \sum_\m \fo_\m\, \zetao(t,r)\, Y^{3,\m}_A 
- \frac{1}{2} \Omega \sum_\m \zetahatq(t,r)\, 
\partial_\phi X^{2,\m}_A, 
\label{zetaA_spharm} \\  
w_A &= \Omega \sum_\m \fd_\m\, \wwd(t,r)\, Y^{1,\m}_A 
+ \frac{1}{3} \Omega \sum_\m \fo_\m\, \wwo(t,r)\, Y^{3,\m}_A 
- \frac{1}{2} \Omega \sum_\m \whatq(t,r)\, 
\partial_\phi X^{2,\m}_A. 
\label{wA_spharm} 
\end{align} 
\end{subequations} 
A decomposition for $\delta \rho$ is not required, because the equation
of state provides a direct relation to $\delta p$. Factors of $\fd_\m$ 
and $\fo_\m$ are inserted within the sums over $\m$ to simplify the
resulting equations, and $\m$ labels on the various coefficients $\pd,
\cdots, \whatq$ are omitted to keep the notation uncluttered. As we
shall see, the perturbation equations satisfied by these quantities
will all be independent of $\m$, except for the driving terms
involving the gravitomagnetic tidal moments $\Bq_\m$. The
infrastructure put in place here, elaborate though it may seem,
produces a substantial simplification of the resulting perturbation
equations.     

We next transform Eqs.~(\ref{euler_vs_zeta1}), (\ref{euler_vs_zeta2}),
and (\ref{Eulerpert3}) from the Cartesian coordinates $x^a$ to the
spherical coordinates $(r,\theta^A)$, and substitute the
decompositions in spherical harmonics. This returns a large
set of equations, with subsets that decouple from one another. In the
dipole sector we have  
\begin{subequations} 
\label{dipole_sector} 
\begin{align} 
& 0 = \partial_{tt} \zetad_r + \partial_r \bigl( \pd/\rho - \Ud \bigr)  
+ \frac{2}{5c^2} r^2 \Bq_\m(t), 
\label{Er_d} \\
& 0 = \partial_{tt} \zetad + \pd/\rho - \Ud 
+ \frac{4}{5c^2} r^3 \Bq_\m(t), 
\label{E_d} \\ 
& 0 = \rho \partial_r \zetad_r + (\rho' + 2\rho/r) \zetad_r 
- \frac{2\rho}{r^2} \zetad - \frac{\rho' r^2}{G m} (\pd/\rho), 
\label{L_d} \\
& 0 = \partial_{rr} \Ud + \frac{2}{r} \partial_r \Ud 
- \frac{2}{r^2} \Ud - \frac{4\pi r^2 \rho'}{m} (\pd/\rho), 
\label{P_d} \\ 
& \wwd_r = \partial_t \zetad_r, \qquad 
\wwd = \partial_t \zetad, 
\label{w_d} 
\end{align} 
\end{subequations} 
where $\rho' := d\rho/dr$. In the octupole sector we have 
\begin{subequations} 
\label{octupole_sector} 
\begin{align} 
& 0 = \partial_{tt} \zetao_r + \partial_r \bigl( \po/\rho - \Uo \bigr)  
- \frac{2}{3c^2} r^2 \Bq_\m(t), 
\label{Er_o} \\
& 0 = \partial_{tt} \zetao + 3\bigl( \po/\rho - \Uo \bigr)  
- \frac{2}{3c^2} r^3 \Bq_\m(t), 
\label{E_o} \\ 
& 0 = \rho \partial_r \zetao_r + (\rho' + 2\rho/r) \zetao_r 
- \frac{4\rho}{r^2} \zetao - \frac{\rho' r^2}{G m} (\po/\rho), 
\label{L_o} \\
& 0 = \partial_{rr} \Uo + \frac{2}{r} \partial_r \Uo 
- \frac{12}{r^2} \Uo - \frac{4\pi r^2 \rho'}{m} (\po/\rho), 
\label{P_o} \\ 
& \wwo_r = \partial_t \zetao_r, \qquad 
\wwo = \partial_t \zetao, 
\label{w_o} 
\end{align} 
\end{subequations} 
and the quadrupole sector is limited to 
\begin{subequations}
\label{quadrupole_sector}  
\begin{align} 
& 0 = \partial_{tt} \zetahatq + \frac{4}{9c^2} r^3 \Bq_\m(t), 
\label{E_q} \\ 
& \whatq = \partial_t \zetahatq 
+ \frac{2}{3c^2} r^3 \int^t \Bq_\m(t')\, dt'. 
\label{w_q} 
\end{align} 
\end{subequations} 
 
\section{Solution to the perturbation equations}
\label{sec:solution} 

In this section we integrate the perturbation equations displayed in
Eqs.~(\ref{dipole_sector}), (\ref{octupole_sector}), and
(\ref{quadrupole_sector}). For concreteness and simplicity we choose a
stellar model corresponding to the polytropic equation of state 
$p = K \rho^2$, where $K$ is a constant. The structure equations for
this model return  
\begin{equation} 
\rho = \frac{M}{4R^2 r} \sin(\pi r /R), \qquad 
m = \frac{M}{\pi} \biggl[ \sin(\pi r/R) 
- \frac{\pi r}{R} \cos(\pi r/R) \biggr]
\label{polytrope1} 
\end{equation} 
for the density and mass functions, respectively, and 
\begin{equation} 
p = \frac{GM^2}{8\pi R^2 r^2} \sin^2(\pi r/R) 
\label{polytrope2} 
\end{equation} 
for the pressure. The equations also return $K = 2GR^2/\pi$ for the
constant appearing in the equation of state. 

\subsection{Quadrupole sector} 
\label{subsec:quadrupole} 

The solution to Eq.~(\ref{E_q}) is immediate, and actually independent
of the equation of state: 
\begin{equation} 
\zetahatq = -\frac{4}{9c^2} r^3 
\int^t_{-\infty} (t-t') \Bq_\m(t')\, dt'.  
\end{equation} 
Equation (\ref{w_q}) then gives 
\begin{equation} 
\whatq = \frac{2}{9c^2} r^3 \int^t_{-\infty} 
\Bq_\m(t')\, dt',  
\end{equation} 
and this represents a growing contribution to the velocity
field. Substituting this within Eq.~(\ref{wA_spharm}) and recalling
Eq.~(\ref{Bhatq_pot}), we find that the quadrupole term in the
velocity perturbation is given by   
\begin{equation} 
w_A^{\ell=2} = \frac{2}{9c^2} r^3 \int^t_{-\infty} 
\Bhatq_A(t')\, dt'; 
\label{wA_quad} 
\end{equation} 
the radial component of the velocity field vanishes. 

The velocity perturbation becomes 
\begin{equation} 
w_a^{\ell=2} = \frac{2}{9c^2} \epsilon_{abc} x^b x^d 
\int^t_{-\infty} \hat{\B}^c_{\ d}(t')\, dt' 
\label{wa_quadrupole} 
\end{equation} 
after conversion to Cartesian coordinates, with $\hat{\B}_{ab}$
defined by Eq.~(\ref{Bhat_def}). For the specific tidal environment
described by Eq.~(\ref{Bab_circular}) and corresponding to a companion
body of mass $M'$ moving on a circular orbit of radius $r'$ in the
body's equatorial plane, the quadrupole velocity field is 
\begin{subequations} 
\begin{align} 
w^{\ell=2}_x &= -\frac{G M'  v' R^2}{3 c^2 r^{\prime 3}} 
\frac{\Omega}{\omega'} \Bigl[ (3\cos^2\theta-1) \sin\Phi 
+ \sin^2\theta \sin(\Phi-2\phi) \Bigr] \bar{r}^2, \\ 
w^{\ell=2}_y &= \frac{G M'  v' R^2}{3 c^2 r^{\prime 3}} 
\frac{\Omega}{\omega'} \Bigl[ (3\cos^2\theta-1) \cos\Phi 
- \sin^2\theta \cos(\Phi-2\phi) \Bigr] \bar{r}^2, \\ 
w^{\ell=2}_z &= \frac{2 G M'  v' R^2}{3 c^2 r^{\prime 3}} 
\frac{\Omega}{\omega'} \sin\theta\cos\theta  
\sin(\Phi-\phi)\, \bar{r}^2,
\end{align}
\end{subequations} 
where $v' = r'\omega'$ is the orbital velocity, $\omega'$ the orbital
angular velocity of Eq.~(\ref{orbital_omega}), $\Phi := \omega' t$ the
orbital phase, $(\theta,\phi)$ the polar angles associated with the
coordinates $x^a$, and $\bar{r} := r/R$. To arrive at these
expressions it was assumed that $r'$ varies over a radiation-reaction
time scale that is much longer than $1/\omega'$, and that 
$r' = \infty$ at  $t = -\infty$; this is consistent with our previous
assumption that the body begins in an unperturbed state.     

\subsection{Octupole sector} 
\label{subsec:octupole} 

Combining Eqs.~(\ref{Er_o}) and (\ref{E_o}) yields 
$\partial_{tt}( \zetao_r - \frac{1}{3} \partial_r \zetao ) = 0$, with
the terms involving the driving force cancelling out. The vanishing
initial conditions at $t=-\infty$ imply that 
$\zetao_r - \frac{1}{3} \partial_r \zetao = 0$ at all times, and this
combination of perturbation quantities is therefore unable to grow in
time. While $\zetao_r$ and $\zetao$ could grow individually, we assume
that this does not occur. In view of Eq.~(\ref{zeta_grow}), this
amounts to an assumption that the $\ell=3$ piece of $B_a^{\rm zf}$
actually vanishes, which prevents the growth of the octupole piece
of the Lagrangian displacement vector. This assumption will be
justified in Sec.~\ref{sec:ZFmodes}.  

To reflect this assumption we make the ansatz 
\begin{equation} 
\zetao_r = \yo_r(r)\, \Bq_\m(t), \qquad 
\zetao = \yo(r)\, \Bq_\m(t)  
\end{equation} 
for the Lagrangian displacement, with $\yo_r 
= \frac{1}{3} d\yo/dr$, and we write 
\begin{equation} 
\po = \ppo(r)\, \Bq_\m(t), \qquad 
\Uo = \UUo(r)\, \Bq_\m(t) 
\end{equation} 
for the remaining perturbations. We neglect the time derivatives when
we make the substitutions into the perturbation equations. Equation
(\ref{E_o}) then produces 
\begin{equation} 
\ppo/\rho = \UUo +\frac{2}{9c^2} r^3, 
\label{po_sol1} 
\end{equation}
and inserting this in Eq.~(\ref{P_o}) yields 
\begin{equation} 
\frac{ d^2 \UUo}{dr^2} + \frac{2}{r} \frac{d \UUo}{dr} 
- \biggl( \frac{12}{r^2} + \frac{4\pi r^2 \rho'}{m} \biggr) \UUo 
= \frac{2}{9c^2} \frac{4\pi r^5 \rho'}{m}. 
\label{Uo_eq}  
\end{equation} 
Equation (\ref{L_o}) gives rise to a second-order differential
equation for $\yo$, which we shall not concern ourselves with,
since the physical aspects of the perturbation are completely captured
by Eqs.~(\ref{po_sol1}) and (\ref{Uo_eq}).  

Equation (\ref{Uo_eq}) can be solved analytically for the polytropic
model introduced previously. We require the solution to be regular at
$r=0$, and to match smoothly with an external solution of the form 
$\UUo \propto r^{-4}$ at $r=R$. This solution is given by 
\begin{equation} 
\UUo = -\frac{2}{9c^2} r^3 \biggl[ 1 
+ \frac{7 R^5}{\pi^3 r^7} \bigl( 2\pi^2 r^2 - 5 R^2 \bigr) 
\sin(\pi r/R)
- \frac{7 R^4}{3\pi^2 r^6} \bigl( \pi^2 r^2 - 15 R^2 \bigr) 
\cos(\pi r/R) \biggr]. 
\label{Uo_sol} 
\end{equation} 
The function within square brackets behaves as $1 - \pi^4/45 
+ O(r^2/R^2)$ close to $r=0$, and as $5(2\pi^2-21)/(3\pi^2) 
+ O(1-r/R)$ close to $r=R$. Equation (\ref{po_sol1}) then gives  
\begin{equation} 
\ppo/\rho = -\frac{2}{9c^2} r^3 \biggl[  
\frac{7 R^5}{\pi^3 r^7} \bigl( 2\pi^2 r^2 - 5 R^2 \bigr) 
\sin(\pi r/R)
- \frac{7 R^4}{3\pi^2 r^6} \bigl( \pi^2 r^2 - 15 R^2 \bigr) 
\cos(\pi r/R) \biggr].  
\label{po_sol2} 
\end{equation} 
In this expression, the function within square brackets behaves as
$-\pi^4/45 + O(r^2/R^2)$ close to $r=0$, and as $7(\pi^2-15)/(3\pi^2)
+ O(1-r/R)$ close to $r=R$. 

It can be observed that $\ppo/\rho$ approaches a nonzero value at 
$r = R$; with $\rho(R) = 0$, this means that $\ppo$ itself
vanishes at the stellar surface. The correct surface condition can be
inferred from $\Delta \rho = 0$ at $r=R$, which implies $\delta \rho  
+ \rho' \zeta_r = 0$. On the other hand, the equation of state and the
structure equations imply $\delta p/\rho = -(G m/r^2)
\delta\rho/\rho'$, and combining these equations yields 
\begin{equation} 
\frac{\delta p}{\rho}  \biggr|_{r=R} = 
\frac{GM}{R^2} \zeta_r(r=R). 
\label{deltap_surface} 
\end{equation} 
Because $\zeta_r \neq 0$ at the surface, it follows that 
$\delta p/\rho$ must be nonvanishing as well.   

Inserting Eqs.~(\ref{Uo_sol}) and (\ref{po_sol2}) into
Eqs.~(\ref{deltaU_spharm}) and (\ref{deltap_spharm}), respectively,
and recalling Eqs.~(\ref{assocs}), (\ref{Kd_Ko}), and (\ref{Ko_pot}), 
we find that the octupole piece of the potential and pressure
perturbations are given by 
\begin{equation} 
\delta U^{\ell = 3} = -\frac{2}{9c^2} \K_{abc} x^a x^b x^c 
\biggl[ 1 
+ \frac{7 R^5}{\pi^3 r^7} \bigl( 2\pi^2 r^2 - 5 R^2 \bigr) 
\sin(\pi r/R)
- \frac{7 R^4}{3\pi^2 r^6} \bigl( \pi^2 r^2 - 15 R^2 \bigr) 
\cos(\pi r/R) \biggr] 
\label{dU_octupole} 
\end{equation} 
and 
\begin{equation} 
\delta p^{\ell = 3}/\rho = -\frac{2}{9c^2} \K_{abc} x^a x^b x^c  
\biggl[ \frac{7 R^5}{\pi^3 r^7} \bigl( 2\pi^2 r^2 - 5 R^2 \bigr) 
\sin(\pi r/R)
- \frac{7 R^4}{3\pi^2 r^6} \bigl( \pi^2 r^2 - 15 R^2 \bigr) 
\cos(\pi r/R) \biggr],  
\end{equation} 
where $\K_{abc}$ is defined by Eq.~(\ref{K_def}). 

Our expression for $\delta U^{\ell=3}$ is used in Appendix
\ref{app:Love} to calculate the rotational-tidal Love number of the 
polytropic stellar model. 

\subsection{Dipole sector} 
\label{subsec:dipole} 

Combining Eqs.~(\ref{Er_d}) and (\ref{E_d}) gives 
\begin{equation} 
\partial_{tt} \bigl( \zetad_r - \partial_r \zetad \bigr) 
= \frac{2}{c^2} r^2 \Bq_\m(t), 
\end{equation} 
and in this case we see a nonzero driving force on the right-hand side
of the equation. This implies that $\zetad_r - \partial_r \zetad$ must
grow with time, and to proceed we assume that $\zetad_r$ and $\zetad$
grow individually. To reflect this we make the ansatz   
\begin{subequations} 
\label{zeta_vs_z} 
\begin{align} 
\zetad_r &= \zd_r(r) \int_{-\infty}^t (t-t') \Bq_\m(t')\, dt'   
+ \yd_r(r)\, \Bq_\m(t), \\
\zeta_r &= \zd(r) \int_{-\infty}^t (t-t') \Bq_\m(t')\, dt'
+ \yd(r)\, \Bq_\m(t) 
\end{align}
\end{subequations}  
for the Lagrangian displacement vector, and 
\begin{equation} 
\pd = \ppd(r)\, \Bq_\m(t), \qquad 
\Ud = \UUd(r)\, \Bq_\m(t) 
\end{equation} 
for the remaining perturbations. The assumption leading to
Eq.~(\ref{zeta_vs_z}) will be justified in Sec.~\ref{sec:ZFmodes},
where we show that the dipole piece of $B_a^{\rm zf}$ does not vanish
and therefore leads to a growing displacement vector. 

We insert the preceding equations into Eqs.~(\ref{dipole_sector}),
neglect terms involving second derivatives of $\Bq_\m(t)$, and obtain
the system of equations  
\begin{subequations} 
\begin{align} 
0 &= \zd_r - \frac{d \zd}{dr} - \frac{2}{c^2} r^2, 
\label{dip1_a} \\ 
0 &= \zd + \ppd/\rho - \UUd + \frac{4}{5c^2} r^3,
\label{dip1_b} \\ 
0 &= \rho \frac{d\zd_r}{dr} + (\rho' + 2\rho/r) \zd_r 
- \frac{2\rho}{r^2} \zd, 
\label{dip1_c} \\ 
0 &= \frac{d^2 \UUd}{dr^2} + \frac{2}{r} \frac{d \UUd}{dr} 
- \frac{2}{r} \UUd - \frac{4\pi r^2 \rho'}{m} (\ppd/\rho) 
\label{dip1_d} 
\end{align} 
\end{subequations} 
for the radial functions $\zd_r$, $\zd$, $\ppd$, and $\UUd$. An
equation can also derived for $\yo_r$ and $\yo$, but these variables
are of no concern to us. 

Equation (\ref{dip1_b}) allows us to eliminate $\zd$ from the system,
and substitution into Eqs.~(\ref{dip1_a}) and (\ref{dip1_c}) produces
\begin{subequations} 
\begin{align} 
0 &= r \frac{d \zd_r}{dr} + \frac{r\rho'+2\rho}{\rho} \zd_r 
+ \frac{2}{r} (\ppd/\rho - \UUd) + \frac{8}{5c^2} r^2, \\ 
0 &= r \frac{d}{dr} (\ppd/\rho) + r \zd_r - r \frac{d\UUd}{dr} 
+ \frac{2}{5c^2} r^3.
\end{align} 
\end{subequations} 
These equations, together with Eq.~(\ref{dip1_d}), form a closed set
of equations for $\zd_r$, $\ppd$, and $\UUd$. These must be integrated
numerically.  

To facilitate the numerical work we introduce the new variables 
$e_1, \cdots, e_4$ defined by   
\begin{equation} 
\zd_r = \frac{R^2}{c^2} e_1, \qquad 
\ppd/\rho = \frac{R^2}{c^2} r e_2, \qquad 
\UUd = \frac{R^2}{c^2} r e_3, \qquad 
\frac{d\UUd}{dr} = \frac{R^2}{c^2} e_4. 
\label{en_def} 
\end{equation} 
We further define the dimensionless radial variable $\bar{r} := r/R$,
density function $\bar{\rho} := (R^3/M) \rho$, and mass function
$\bar{m} := m/M$. In terms of all this we have that 
\begin{equation} 
\zd = \frac{R^2}{c^2} r e_5, \qquad 
e_5 := -e_2 + e_3 - \frac{4}{5} \bar{r}^2,  
\label{e5_def} 
\end{equation} 
and the perturbation equations become  
\begin{subequations} 
\label{e1e4_diffeq} 
\begin{align} 
\bar{r} e_1' &= - \frac{\bar{r} \bar{\rho}' + 2 \bar{\rho}}{\bar{\rho}} e_1 
- 2e_2 + 2e_3 - \frac{8}{5} \bar{r}^2, 
\label{e1_diffeq} \\ 
\bar{r} e_2' &= -e_1 - e_2 + e_4 - \frac{2}{5} \bar{r}^2, 
\label{e2_diffeq} \\ 
\bar{r} e_3' &= -e_3 + e_4, 
\label{e3_diffeq} \\ 
\bar{r} e_4' &= \frac{4\pi \bar{r}^4 \bar{\rho}'}{\bar{m}} e_2 
+ 2e_3 - 2e_4, 
\label{e4_diffeq} 
\end{align} 
\end{subequations} 
in which a prime now indicates differentiation with respect to
$\bar{r}$. 

An examination of Eqs.~(\ref{e1e4_diffeq}) near $\bar{r}=0$ indicates
that the functions $e_n$ all tend to a nonvanishing constant at
$\bar{r}=0$, and that they admit an expansion in powers of
$\bar{r}^2$. Furthermore, the local analysis reveals that $e_1(0)$ and
$e_3(0)$ are freely specifiable constants that determine all other
coefficients in the power expansions. On the other hand, an
examination of the equations near $\bar{r}=1$ shows that except for
$e_1$, all functions tend to a nonvanishing constant at $\bar{r}=1$;
for $e_1$ we must impose $e_1(1)=0$ to account for the singular factor
$\bar{\rho}^{-1}$ in Eq.~(\ref{e1_diffeq}). All functions admit an
expansion in powers of $1-\bar{r}$. We also find that $e'_1(1)$,
$e_3(1)$, and $e_4(1)$ are freely specifiable and determine all other
coefficients in the power expansion. 

A boundary condition at $\bar{r}=1$ is required to make
the problem well posed. As we discuss in Appendix \ref{app:dipole},
the external solution for $\UUd$ must be linear in $\bar{r}$, so that
$e^{\rm ext}_3 = \mbox{constant}$. Equation (\ref{e3_diffeq}) further
implies that $e^{\rm ext}_4 = e^{\rm ext}_3$, and the required surface 
condition is therefore that $e_4(1) = e_3(1)$. With this we have a
total of four boundary values that cannot be determined solely from a
local analysis near $\bar{r}=0$ and $\bar{r}=1$; these are $e_1(0)$,
$e_3(0)$, $e'_1(1)$, and $e_3(1)$. A global integration is required to
determine all four constants, and a practical approach is to shoot
toward a middle point: We first integrate the equations from
$\bar{r}=0$ up to a middle point $\bar{r} = \bar{r}_1$, then integrate
them again from $\bar{r}=1$ down to $\bar{r}_1$, and search for the
boundary values that force the two sets of solutions to agree with
each other at $\bar{r}=\bar{r}_1$. A concrete implementation of this
method is described in Sec.~17.2 of {\it Numerical Recipes}
\cite{numerical-recipes:02}.   

\begin{figure} 
\includegraphics[width=0.9\linewidth]{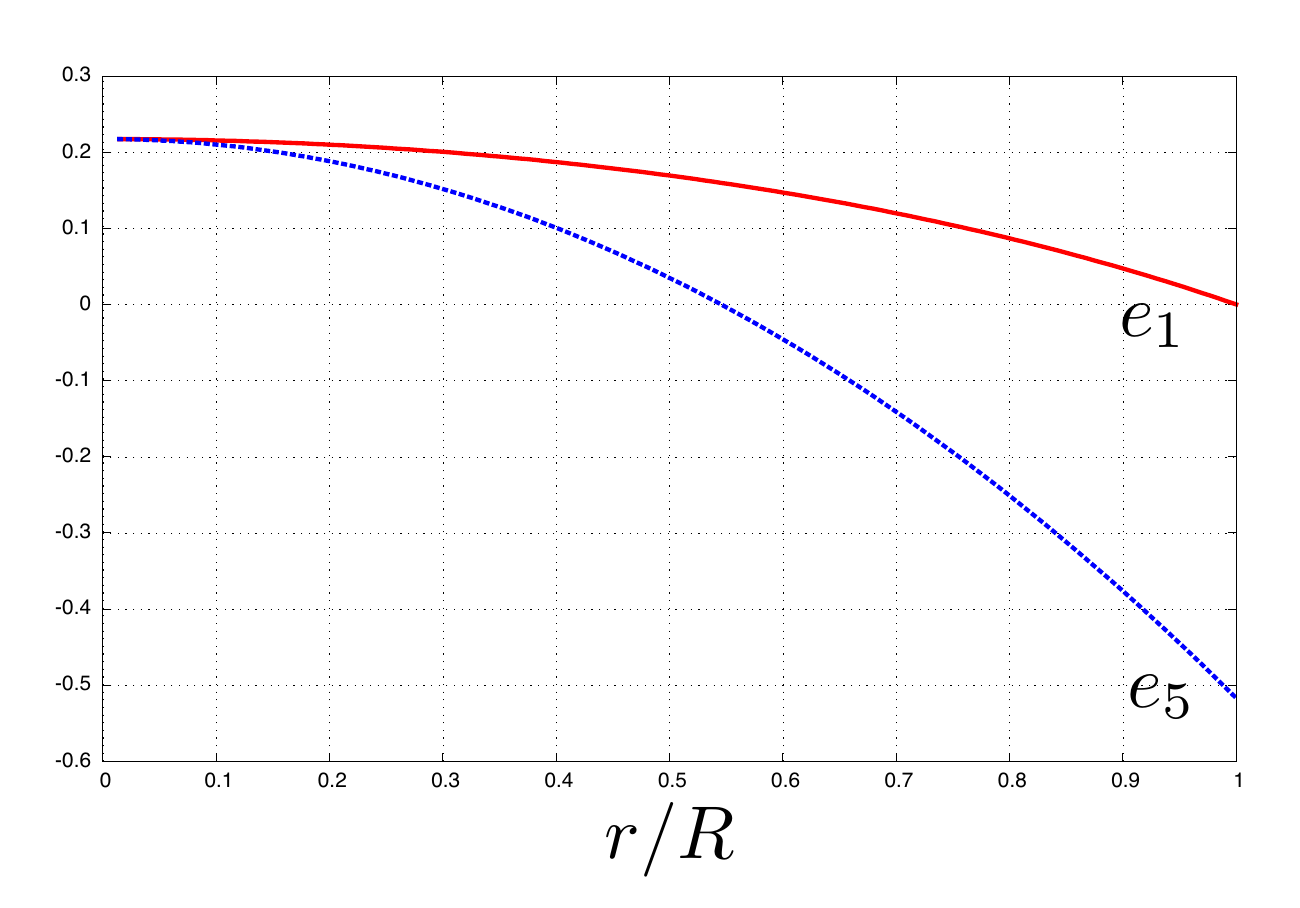}
\caption{Numerical solution for the functions $e_1$ and $e_5$ that
  substitute for the variables $\zd_r$ and $\zd$, respectively,
  plotted as functions of $\bar{r} = r/R$. The numerical error is well
  within the thickness of the curves.} 
\label{fig:velocity} 
\end{figure} 

\begin{figure} 
\includegraphics[width=0.9\linewidth]{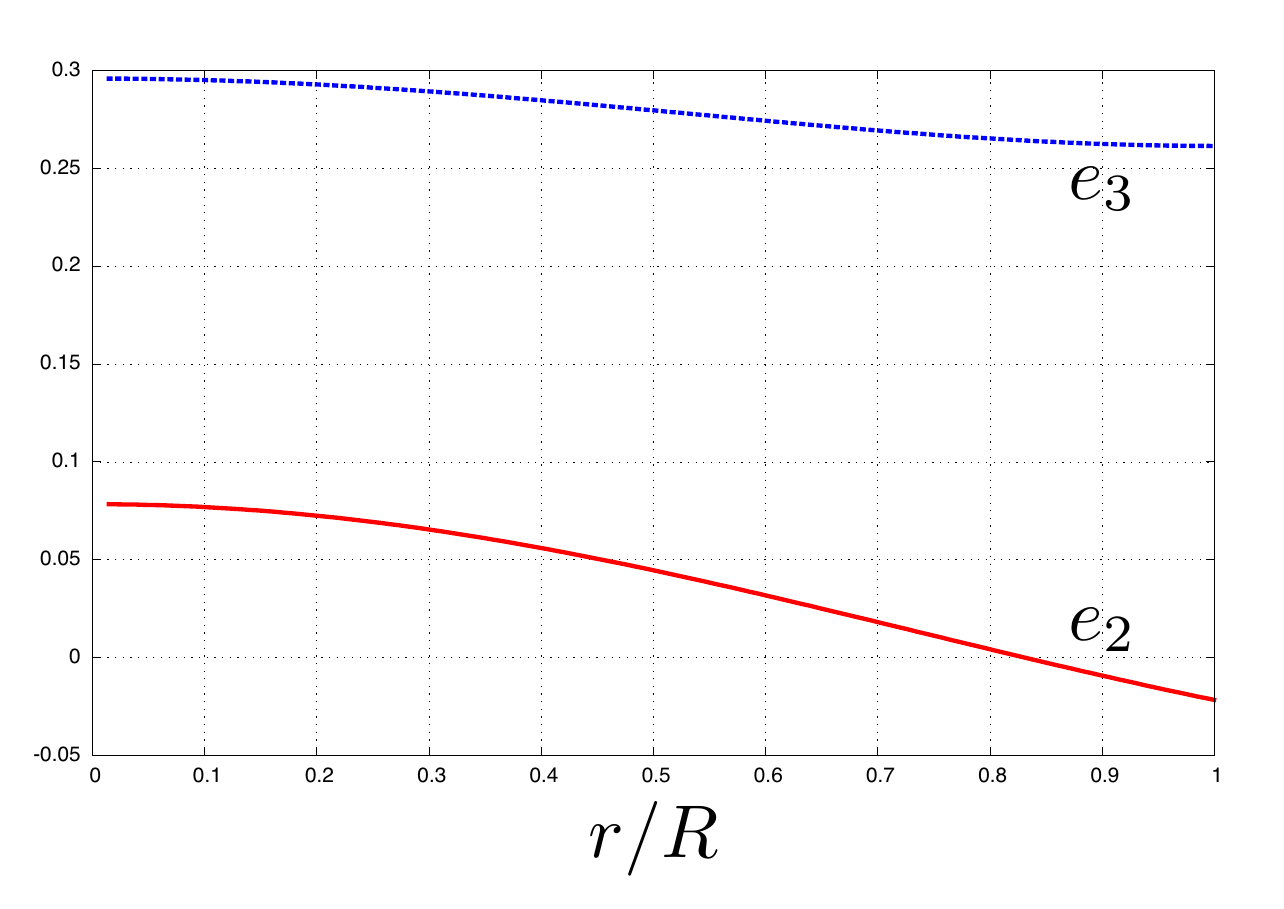}
\caption{Numerical solution for the functions $e_2$ and $e_3$ that
  substitute for the variables $\ppd$ and $\UUd$, respectively,
  plotted as functions of $\bar{r} = r/R$. The numerical error is well
  within the thickness of the curves.} 
\label{fig:pressure_potential} 
\end{figure} 

For the numerical work we adopt the polytropic model described by
Eqs.~(\ref{polytrope1}) and (\ref{polytrope2}). For this specific case
we have 
\begin{equation} 
\frac{\bar{r} \bar{\rho}' + 2 \bar{\rho}}{\bar{\rho}} 
= \frac{\sin(\pi\bar{r}) + (\pi\bar{r})\cos(\pi\bar{r})}
  {\sin(\pi\bar{r})}, \qquad 
\frac{4\pi \bar{r}^4 \bar{\rho}'}{\bar{m}}  
= -(\pi \bar{r})^2. 
\end{equation} 
The numerical solutions for $e_1$, $e_2$, $e_3$, and $e_5$ are
displayed in Figs.~\ref{fig:velocity} and
\ref{fig:pressure_potential}; the solution for $e_4$ is not
shown, because it can be obtained directly from $e_3$ by
exploiting Eq.~(\ref{e3_diffeq}). 

According to Eqs.~(\ref{w_d}) and (\ref{zeta_vs_z}), we have that 
\begin{equation} 
\wwd_r = \zd_r \int_{-\infty}^t \Bq_\m(t')\, dt', \qquad 
\wwd = \zd \int_{-\infty}^t \Bq_\m(t')\, dt',
\end{equation} 
in which we neglected terms proportional to $\partial_t
\Bq_\m$. Substituting this within Eqs.~(\ref{wr_spharm}) and
(\ref{wA_spharm}) and recalling Eqs.~(\ref{Kd_Ko}) and (\ref{Kd_pot}), 
we find that the dipole piece of the velocity perturbation is given by  
\begin{equation} 
w_r^{\ell=1} = \zd_r \int_{-\infty}^t \Kd(t')\, dt', \qquad 
w_A^{\ell=1} = \zd \int_{-\infty}^t \Kd_A(t')\, dt'. 
\label{wrA_dipole} 
\end{equation} 
If we next express this in terms of the radial functions $e_1$ and
$e_5$ by invoking Eqs.~(\ref{en_def}) and (\ref{e5_def}), and convert
to Cartesian coordinates, we get 
\begin{equation} 
w_a^{\ell=1} = \frac{R^2}{c^2} \bigl[ 
e_1\, n_a n^b + e_5\, ( \delta_a^{\ b} - n_a n^b ) \bigr] 
\int_{-\infty}^t \K_b(t')\, dt',
\label{wa_dipole} 
\end{equation} 
where $\K_a$ is defined by Eq.~(\ref{K_def}). For the specific tidal
environment provided by a companion body of mass $M'$ moving on a
circular orbit of radius $r'$ in the body's equatorial plane, the
quadrupole velocity field is  
\begin{subequations} 
\begin{align} 
w^{\ell=1}_x &= \frac{3G M'  v' R^2}{2 c^2 r^{\prime 3}} 
\frac{\Omega}{\omega'} \Bigl\{ \sin^2\theta \bigl[ \sin\Phi 
+ \sin(\Phi-2\phi) \bigr] e_1 + \bigl[ (\cos^2\theta+1) \sin\Phi 
- \sin^2\theta \sin(\Phi - 2\phi) \bigr] e_5 \Bigr\}, \\ 
w^{\ell=1}_y &= -\frac{3G M'  v' R^2}{2 c^2 r^{\prime 3}} 
\frac{\Omega}{\omega'} \Bigl\{ \sin^2\theta \bigl[ \cos\Phi 
- \cos(\Phi-2\phi) \bigr] e_1 + \bigl[ (\cos^2\theta+1) \cos\Phi 
+ \sin^2\theta \sin(\Phi - 2\phi) \bigr] e_5 \Bigr\}, \\ 
w^{\ell=1}_z &= \frac{3 G M'  v' R^2}{c^2 r^{\prime 3}} 
\frac{\Omega}{\omega'} \sin\theta\cos\theta  
\sin(\Phi-\phi)\, (e_1-e_5),
\end{align}
\end{subequations} 
where $v' = r'\omega'$ is the orbital velocity, $\omega'$ the orbital
angular velocity of Eq.~(\ref{orbital_omega}), $\Phi := \omega' t$ the
orbital phase, and $(\theta,\phi)$ the polar angles associated with
the coordinates $x^a$. 

A calculation similar to the one leading to Eq.~(\ref{wa_dipole})
reveals that the dipole piece of the pressure and potential 
perturbations are given by  
\begin{equation} 
\delta p^{\ell=1}/\rho = \frac{R^2}{c^2} e_2\, \K_a x^a, \qquad 
\delta U^{\ell=1} = \frac{R^2}{c^2} e_3\, \K_a x^a. 
\label{dU_dipole} 
\end{equation} 
In Appendix \ref{app:CM} we construct the dipole piece of the
acceleration field, and verify that its mass-weighted average gives a
vanishing acceleration for the body's center-of-mass.   

\section{Zero-frequency modes} 
\label{sec:ZFmodes} 

In this section we examine the zero-frequency modes $f_I^a$
introduced in Sec.~\ref{sec:mode}, and show that they are directly
responsible for the velocity fields displayed in
Eqs.~(\ref{wa_quadrupole}) and (\ref{wa_dipole}). The completion of
the mode analysis initiated in Sec.~\ref{sec:mode} provides a complete
justification of the assumptions made in Sec.~\ref{sec:solution}
concerning the form of solution to the perturbation equations in the
dipole and octupole sectors. As in the rest of the paper we assume
that the fluid is barotropic, with the perturbed fluid possessing the
same equation of state as the unperturbed fluid. In this case it is
known (see Ref.~\cite{lockitch-andersson-friedman:00} for a clear
presentation) that the zero-frequency modes separate into even-parity
$g$-modes and odd-parity $r$-modes. (The $g$-modes do not exist when
the perturbed fluid possesses a distinct equation of state.)   

\subsection{Mode equation} 

Returning to the notation introduced in Sec.~\ref{sec:mode}, the
zero-frequency modes satisfy $\LL_a^{\ b} f_b \equiv -P_a = 0$, which
takes the explicit form displayed in Eq.~(\ref{P_def}). Writing the
equation in spherical coordinates $(r,\theta,\phi)$, we see that the
angular components reduce to $\delta p - \rho \delta U = 0$, and that
the radial component, simplified with the structure equations
(\ref{hydro_eq}), merely reproduces $\delta p = (dp/d\rho) 
\delta \rho$. Inserting these relations into Eq~(\ref{Poisson})
produces $\nabla^2 \delta U - (4\pi r^2 \rho'/m) \delta U = 0$, and it
is not difficult to show that the general solution to this equation
cannot be smoothly matched to an external solution that is required to
decay with increasing $r$. One way to establish this is to perform a
decomposition in spherical harmonics, observe that each 
$\delta U_{\ell m}(r)$ satisfies a homogeneous equation, that the
solution regular at $r=0$ comes with a single integration constant (an
overall multiplicative factor), and that this single constant is
insufficient to match both $\delta U_{\ell m}(r)$ and its first
derivative to the external solution 
$\delta U_{\ell m} \propto r^{-\ell+1}$ at $r=R$.  

The conclusion is that the zero-frequency modes describe a
perturbation with $\delta \rho = \delta p = \delta U = 0$.  
Equation (\ref{euler_vs_zeta2}) then implies that the mode functions
are constrained by   
\begin{equation} 
\partial_a (\rho f^a) = 0.
\label{zf_mode} 
\end{equation} 
A displacement vector $\zeta_a = f_a$ would describe an entirely
trivial perturbation with vanishing $\delta \rho$, $\delta p$, 
$\delta U$, and $\delta v_a$. But the displacement vector 
$\zeta_a = f_a t$ also satisfies $\partial_{tt} \zeta_a - P_a = 0$,
and it does give rise to a nontrivial velocity field $\delta v_a
= \partial_t \zeta_a = f_a$. A zero-frequency perturbation is
therefore a velocity field constrained by Eq.~(\ref{zf_mode}). 

\subsection{Basis of zero-frequency modes}  

We transform Eq.~(\ref{zf_mode}) to spherical coordinates
$(r,\theta^A)$ and consider solutions of the factorized form 
\begin{equation} 
f_r = f_r^{\ell\m}\, Y^{\ell\m}, \qquad 
f_A = f^{\ell\m}\, Y^{\ell\m}_A  
\end{equation} 
for the even-parity $g$-modes, and 
\begin{equation}
f_r = 0, \qquad  
f_A = \hat{f}^{\ell\m}\, X^{\ell\m}_A 
\end{equation} 
for the odd-parity $r$-modes, where $Y^{\ell\m}$, $Y^{\ell\m}_A$, and
$X^{\ell\m}_A$ are the spherical harmonics introduced in
Sec.~\ref{sec:harmonics}; the functions $f_r^{\ell\m}$, $f^{\ell\m}$,
and $\hat{f}^{\ell\m}$ depend on $r$ only. Making the substitutions
in Eq.~(\ref{zf_mode}) reveals that the $g$-mode functions are
constrained by 
\begin{equation} 
\ell(\ell+1) \rho f^{\ell\m} 
= \frac{d}{dr} \bigl( r^2 \rho f_r^{\ell\m} \bigr), 
\label{gmode} 
\end{equation} 
so that $f^{\ell\m}$ is determined once $f^{\ell\m}_r$ is
specified. The exercise further reveals that the $r$-mode function is
completely unconstrained. The zero-frequency modes are therefore
characterized by two freely specifiable functions, $f_r^{\ell\m}$ and
$\hat{f}^{\ell\m}$. We have two infinitely degenerate sets of modes. 

Two $g$-modes, $\bm{a}$ and $\bm{b}$, which share the same values of
$\ell$ and $\m$, have a scalar product defined by  
\begin{equation} 
\langle \bm{a}, \bm{b}\rangle := \int \rho\, \bm{a} \cdot \bm{b}\, d^3x 
= N^{\ell\m} \biggl[ \int_0^R \rho\, a_r^{\ell\m} b_r^{\ell\m} r^2\, dr 
+ \ell(\ell+1) \int_0^R \rho\, a^{\ell\m} b^{\ell\m}\, dr \biggr], 
\label{gmode_scalar} 
\end{equation} 
where $N^{\ell\m} := \int (Y^{\ell\m})^2\, \sin\theta\, d\theta
d\phi$; modes with different values of either $\ell$ or $\m$ are
orthogonal. Similarly, $r$-modes $\bm{p}$ and $\bm{q}$ have the 
scalar product  
\begin{equation} 
\langle \bm{p}, \bm{q}\rangle = \ell(\ell+1) N^{\ell\m} 
\int \rho\, \hat{p}^{\ell\m} \hat{q}^{\ell\m}\, dr 
\end{equation} 
when they share the same values of $\ell$ and $\m$. All $g$-modes are 
orthogonal to all $r$-modes. 

We wish to construct a basis of orthogonal modes, labelled by $k = 0, 
1, 2, \cdots$ in addition to the spherical-harmonic labels $\ell\m$;
the complete mode label is therefore $I := \ell\m k$. The procedure is
simple, and we describe it in detail in the case of $g$-modes. We
begin with a set of seed modes $\bm{a}^k$ characterized by a
freely-specified $a_r^{\ell\m k}$ and an $a^{\ell\m k}$ determined by 
Eq.~(\ref{gmode}). These modes are not mutually orthogonal, but they
can be turned into a set of orthogonal modes by implementing a
Gram-Schmidt procedure. We first set $\bm{f}^0 = \bm{a}^0$, and then
set 
\begin{equation}
f_r^{\ell\m k} = a_r^{\ell\m k} - \sum_{n=0}^{k-1} 
\frac{ \langle \bm{f}^n, \bm{a}^k \rangle }
        { \langle \bm{f}^n, \bm{f}^n \rangle } a_r^{\ell\m n}  
\end{equation} 
for each successive $k$, with $f^{\ell\m k}$ determined at each stage
by Eq.~(\ref{gmode}). 

The overlap integrals of Eq.~(\ref{mode_amplitudes}) can now be
evaluated. Again we describe the procedure in detail in the case of
$g$-modes. The components of the external force $\bm{B}$ are expanded
in spherical harmonics according to 
\begin{equation} 
B_r = \sum_{\ell\m} B_r^{\ell\m} Y^{\ell\m}, \qquad 
B_A =\sum_{\ell\m} B^{\ell\m} Y_A^{\ell\m}
+ \sum_{\ell\m} \hat{B}^{\ell\m} X_A^{\ell\m},  
\end{equation} 
and Eq.~(\ref{mode_amplitudes}) turns into the explicit form 
\begin{equation} 
B^{\ell\m k} = \frac{N^{\ell\m}}{\langle \bm{f}^k, \bm{f}^k \rangle} 
\biggl[  \int_0^R \rho\, B^{\ell\m}_r f^{\ell\m k} r^2\, dr 
+ \ell(\ell+1) \int_0^R \rho\, B^{\ell\m} f^{\ell\m k}\, dr \biggr] 
\end{equation} 
for the mode amplitudes $B^I$. This can be simplified by inserting
Eq.~(\ref{gmode}) within the second integral and integrating by
parts; we arrive at 
\begin{equation} 
B^{\ell\m k} = \frac{N^{\ell\m}}{\langle \bm{f}^k, \bm{f}^k \rangle} 
\int_0^R \rho\, \biggl( B^{\ell\m}_r - \frac{d B^{\ell\m}}{dr}
\biggr) f_r^{\ell\m k} r^2\, dr. 
\label{gmode_overlap} 
\end{equation} 
With this, the components of $\bm{B}^{\rm zf}$ defined below
Eq.~(\ref{w_grow}) are given by 
\begin{equation} 
B_r^{\rm zf} = \sum_{\ell\m k} B^{\ell\m k}\, f^{\ell\m k}_r\, 
Y^{\ell\m}, \qquad 
B_A^{\rm zf} = \sum_{\ell\m k} B^{\ell\m k}\, f^{\ell\m k}\, 
Y_A^{\ell\m}, 
\end{equation} 
and these are then ready to be inserted within Eq.~(\ref{w_grow}) to 
obtain the growing piece of the velocity perturbation.  

\subsection{Quadrupole sector} 
 
According to Eq.~(\ref{B_decomposed}), the quadrupole piece of the
driving force $\bm{B}$ has the nonvanishing components 
\begin{equation} 
B^{\ell=2}_A = \frac{1}{9} r^3 \Bhatq_A, 
\end{equation} 
and these admit a decomposition in odd-parity harmonics $X_A^{2,\m}$
with coefficients $\hat{B}^{2,\m}$. Because the $r$-modes are
unconstrained, we have the freedom to assign 
$\hat{f}^{2,\m,0} = \hat{B}^{2,\m}$. And because all other members of  
the basis of modes are orthogonal to the zeroth member, we immediately
find that 
\begin{equation} 
\bm{B}^{\ell = 2}_{\rm zf} = \bm{B}^{\ell=2}. 
\end{equation} 
The growing solution of Eq.~(\ref{zeta_grow}) can then be seen to give 
rise to the velocity field of Eq.~(\ref{wA_quad}). In this specific
case the mode analysis is entirely trivial and merely reproduces our
previous results. 

\subsection{Octupole sector} 

Returning to Eq.~(\ref{B_decomposed}), we see that the octupole piece
of the driving force has the components 
\begin{equation} 
B^{\ell=3}_r = -\frac{1}{6} r^2 \Ko, \qquad 
B^{\ell=3}_A = -\frac{1}{6} r^3 \Ko_A, 
\end{equation} 
which admit a decomposition in even-parity harmonics $Y^{3,\m}$ and
$Y^{3,\m}_A$ with coefficients $B^{3,\m}_r = -\frac{1}{6} r^2 \Ko_\m$
and $B^{3,\m} = -\frac{1}{18} r^3 \Ko_\m$, respectively. These are
related by  
\begin{equation} 
B^{3,\m}_r - \frac{d B^{3,\m}}{dr} = 0, 
\end{equation} 
and Eq.~(\ref{gmode_overlap}) implies that all mode amplitudes
$B^{3,\m,k}$ necessarily vanish. In this case we have that 
\begin{equation} 
\bm{B}^{\ell = 3}_{\rm zf} = 0, 
\end{equation} 
and this justifies the assumption made at the beginning of
Sec.~\ref{subsec:octupole}, that the octupole piece of the velocity
field does not possess a growing term. 

\subsection{Dipole sector} 

Returning once more to Eq.~(\ref{B_decomposed}), we see that the
dipole piece of the driving force has the components 
\begin{equation} 
B^{\ell=1}_r = \frac{1}{10} r^2 \Kd, \qquad 
B^{\ell=1}_A = \frac{1}{5} r^3 \Kd_A, 
\end{equation} 
which admit a decomposition in even-parity harmonics with coefficients
$B^{1,\m}_r = \frac{1}{10} r^2 \Kd_\m$ and $B^{1,\m} = \frac{1}{5} r^3
\Kd_\m$, respectively. We choose the mode functions to be independent
of $\m$, denote them $f_r^{k}$ and $f^k$ to simplify the notation
(with the label $\ell = 1$ omitted), and find that in this case, the mode
amplitudes are given by   
\begin{equation} 
B^{1,\m,k} = -\frac{1}{2} \Kd_\m\, \Gamma_{k},
\end{equation} 
where 
\begin{equation} 
\Gamma_{k} := \frac{N^{1,\m}}{\langle \bm{f}^k, \bm{f}^k \rangle}
\int_0^R \rho\, r^4 f^{k}_r\, dr; 
\end{equation} 
these quantities are independent of $\m$ by virtue of the definition
of the scalar product in Eq.~(\ref{gmode_scalar}). With this we have
that  
\begin{equation} 
B^{\rm zf}_r = -\frac{1}{2} \Kd \sum_k \Gamma_k f_r^k, \qquad 
B^{\rm zf}_A = -\frac{1}{2} \Kd_A \sum_k \Gamma_k f^k, 
\end{equation} 
and substitution into Eq.~(\ref{w_grow}) produces the velocity field
of Eq.~(\ref{wrA_dipole}), with $\zd_r = (2/c^2) \sum_k \Gamma_k
f_r^k$ and $\zd = (2/c^2) \sum_k \Gamma_k f^k$. With the definitions
of Eqs.~(\ref{en_def}) and (\ref{e5_def}), this is 
\begin{equation} 
e_1 = \frac{2}{R^2} \sum_k \Gamma_k f_r^k, \qquad 
r e_5 = \frac{2}{R^2} \sum_k \Gamma_k f^k. 
\label{e1_modesum} 
\end{equation} 
At this stage we have justified the assumption made at the beginning
of Sec.~\ref{subsec:dipole}, that the dipole piece of the velocity
field possesses a growing term. In addition, the mode equation
(\ref{gmode}) implies that $e_1$ and $e_5$ are related by 
$2 r \rho\, e_5 = (r^2 \rho\, e_1)'$, with a prime indicating
differentiation with respect to $r$. This relation can also be derived
on the basis of Eq.~(\ref{e1_diffeq}), and we see that the mode
analysis is entirely compatible with the developments of
Sec.~\ref{subsec:dipole}.  

We have yet to verify that the $e_1$ constructed here is precisely
equal to the $e_1$ obtained in Sec.~\ref{subsec:dipole}. For this we
must introduce an actual set of modes $\bm{f}^k$, calculate the
overlap integrals $\Gamma_k$, and carry out the sum over modes. 
To accomplish this we adopt  
\begin{equation} 
a^k_r = \cos \bigl[ {\textstyle \frac{1}{2}} (2k+1) \pi r /R \bigr],
\qquad 
k = 0, 1, 2, \cdots 
\end{equation} 
as a convenient set of seed modes, with the corresponding $a^k$
determined by Eq.~(\ref{gmode}). These mode functions are chosen so
that $a^k_r$ vanishes at $r=R$, as required by the mode equation in
view of the fact that $\rho$ vanishes at the surface, that it tends to
a nonvanishing constant at $r=0$, as required of a dipolar vector
field, and that its derivative with respect to $r$ vanishes at $r=0$,
as expected of $e_1(r)$. We feed the seed modes into the Gram-Schmidt
machine, using the density function of Eq.~(\ref{polytrope1}) to
evaluate the integrals, and obtain a set of orthogonal modes
$\bm{f}^k$. These, finally, are involved in the computation of
$\Gamma_k$ and the mode sum of Eq.~(\ref{e1_modesum}).   

\begin{figure} 
\includegraphics[width=0.9\linewidth]{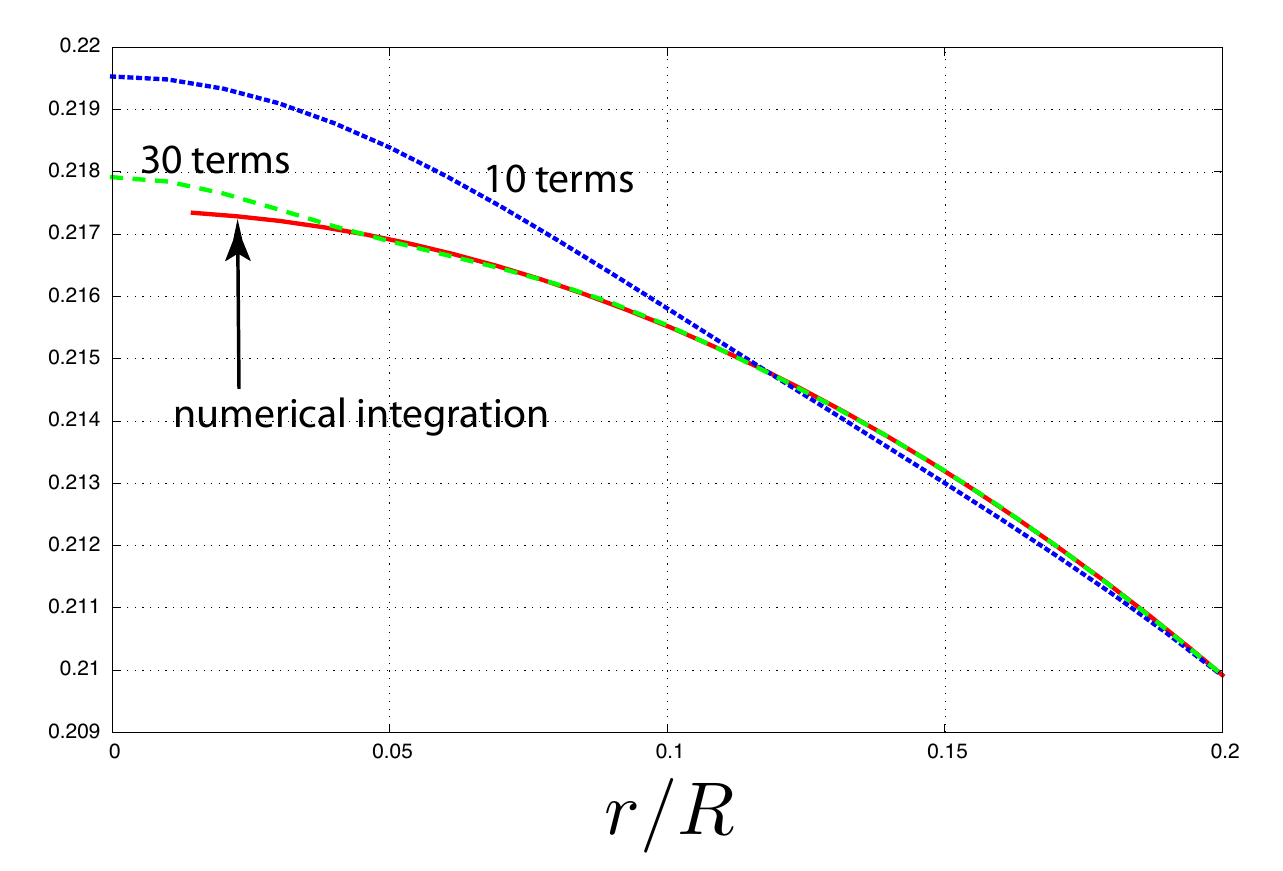}
\caption{Mode-sum representation of $e_1$ compared with the result
  displayed in Fig.~\ref{fig:velocity}. The curve obtained by
  numerical intergation in Sec.~\ref{subsec:dipole} is shown in solid
  red. A construction of $e_1$ involving 10 terms in the mode sum is
  shown in short dashed blue. A construction involving 30 terms is
  shown in long dashed green. The plot focuses on the interval $0 < r/R
  < 0.2$; the curves are indistinguishable beyond this interval.}  
\label{fig:comparison} 
\end{figure} 

At the end of this computation we find that indeed, the $e_1$ of
Eq.~(\ref{e1_modesum}) agrees with the $e_1$ displayed in
Fig.~\ref{fig:velocity}; the comparison is shown in
Fig.~\ref{fig:comparison}. The mode sum converges rapidly when 
$r/R \gtrsim 0.2$; in this range a handful of terms suffice to produce
a curve that is visually identical to the one shown in
Fig.~\ref{fig:velocity}. The sum converges much more slowly when $r$
is small; we find that at least thirty terms are required to
adequately reproduce the curve near the smallest values of $r$
displayed in Fig.~\ref{fig:velocity}.    

\subsection{Conclusion} 

We have shown that the zero-frequency modes are directly responsible 
for the velocity fields displayed in Eqs.~(\ref{wa_quadrupole}) and
(\ref{wa_dipole}). These modes, therefore, play a crucial role in the
gravitomagnetic tidal response of a rotating body. Zero-frequency
modes, however, are typically not involved in the response of a
Newtonian fluid driven by an external force, and they are rarely given
consideration. The reason is that in the typical case, the
external force $B_a$ is the gradient of a scalar potential, 
$B_a = \partial_a \psi$, and this guarantees that the corresponding
mode amplitudes $B_I$ vanish:  
\begin{equation} 
N_I B_I = \int_V \rho\, \bm{f}_I \cdot \bm{\nabla} \psi\, d^3x  
= \oint_S \rho\, \psi \bm{f}_I \cdot d\bm{a} 
- \int_V \psi \bm{\nabla} \cdot (\rho \bm{f}_I)\, d^3x = 0. 
\end{equation} 
The integration domain $V$, bounded by the surface $S$, is chosen to
extend slightly beyond the body, which ensures that $\rho$ always
vanishes on $S$; the second volume integral vanishes by virtue of
Eq.~(\ref{zf_mode}). The zero-frequency modes are involved in the 
gravitomagnetic tidal response of a rotating body because in this
case, the driving force $B_a$ is not a gradient vector field.   

\begin{acknowledgments} 
We thank John Friedman, Phil Landry, Raissa Mendes, and Jean-Philippe
Nicolas for useful conversations, and Raissa Mendes again for a
helpful hand at integrating Eqs.~(\ref{e1e4_diffeq}). One of us (EP)
is grateful for the warm hospitality of the Laboratoire de
Math\'ematiques at the Universit\'e de Bretagne Occidentale, where
part of this work was carried out. The work was supported by the
Natural Sciences and Engineering Research Council of Canada.    
\end{acknowledgments} 

\appendix 
\section{Octupole rotational-tidal Love number} 
\label{app:Love} 

The octupole, rotational-tidal Love number $\KKo$ was introduced in
Secs.~III and IV of Ref.~\cite{landry-poisson:15a}. As explained
there, the Love number provides a (partial) description of the body's
gravitational response to the coupled rotational and gravitomagnetic
tidal perturbations. For our purposes here, the Love number 
is defined by their Eq.~(4.4), in which the external metric of a
slowly rotating, tidally deformed body is presented in Regge-Wheeler
gauge. The relevant term in $g_{tt}$ is 
\begin{equation} 
\delta g_{tt}^{\ell=3} = -\frac{8G}{c^6} \biggl( \frac{2GM}{c^2 r}
\biggr)^4\, \KKo\, S_{\langle a} \B_{bc\rangle} n^a n^b n^c,
\end{equation} 
in which we have replaced $\chi_a$ by $S_a/M^2$, where $S_a$ is the
body's spin angular momentum, restored factors of $G$ and $c$, and
neglected all higher post-Newtonian corrections. This result can be
expressed as  
\begin{equation} 
\delta U_{\rm eff}^{\ell=3} = -\frac{4G}{c^4} 
\biggl( \frac{2GM}{c^2 r} \biggr)^4\, \KKo\, 
S_{\langle a} \B_{bc\rangle} n^a n^b n^c 
\label{dUeff_octupole} 
\end{equation} 
if we introduce an effective gravitational potential via 
$g_{tt} = -1 + 2U_{\rm eff}/c^2$. In our post-Newtonian 
treatment, the body's spin $S^a$ is related to its angular velocity
$\Omega^a$ through the moment of inertia $I$, given by  
\begin{equation} 
I = \frac{8\pi}{3} \int \rho r^4\, dr.  
\end{equation} 
The relation is $S^a = I \Omega^a$, and making the substitution in
Eq.~(\ref{dUeff_octupole}) yields 
\begin{equation} 
\delta U_{\rm eff}^{\ell=3} = 
-\frac{2}{c^2} \biggl( \frac{2GM}{c^2} \biggr)^5
\frac{I}{MR^2}\, \KKo\, \frac{R^2}{r^4}\, \K_{abc} n^a n^b n^c, 
\end{equation} 
with $\K_{abc}$ defined by Eq.~(\ref{K_def}).  

This external expression for the octupole gravitational perturbation
must be matched to the internal expression of
Eq.~(\ref{dU_octupole}) at $r=R$, and this provides the value of
$\KKo$ for the polytropic model considered in this work. In this
case the moment of inertia evaluates to $I/(MR^2) =
2(\pi^2-6)/(3\pi^2)$, and we arrive at 
\begin{equation} 
\KKo = -\kko \biggl(\frac{c^2 R}{2GM}\biggr)^5, \qquad 
\kko = \frac{5}{18} \frac{21-2\pi^2}{\pi^2-6} 
\simeq 9.0505 \times 10^{-2}. 
\end{equation} 
Making the substitution in Eq.~(\ref{dUeff_octupole}) gives 
\begin{equation} 
\delta U_{\rm eff}^{\ell=3} = \frac{2 \kko}{c^2} \frac{R^5}{r^4}\, 
\hat{S}_{\langle a} \B_{bc\rangle} n^a n^b n^c, 
\end{equation} 
where $\hat{S}_a := S_a/M$ is the body's spin per unit mass. The
expression reveals that the body's response to a coupled rotational
and gravitomagnetic tidal perturbation is a post-Newtonian effect that
scales with $R^5$.   

\section{External dipole} 
\label{app:dipole} 

In this Appendix we justify the boundary condition $e_4(1) = e_3(1)$
imposed in Sec.~\ref{subsec:dipole} to integrate the perturbation
equations in the dipole sector. The condition derives from the
statement that in the body's exterior, $\UUd \propto r$: the 
dipole piece of $\delta U$ grows linearly with the distance to the
body's center-of-mass. We note first that the external perturbation
satisfies    
\begin{equation}   
r^2 \frac{d^2 \UUd}{dr^2} + 2r \frac{d \Ud}{dr} - 2\UUd = 0, 
\end{equation} 
with the linearly independent solutions $\UUd \propto r$ and $\UUd
\propto 1/r^2$. Our boundary condition states that we must keep the
growing solution and reject the decaying solution.  

The justification of this statement comes from an examination of the 
external metric of a slowly rotating body subjected to a
gravitomagnetic tidal field. This metric is presented to all
post-Newtonian order in Ref.~\cite{landry-poisson:15a}, and the first  
post-Newtonian approximation of the relevant component $g_{tt}$ is
displayed in Eq.~(8.17a) of Ref.~\cite{poisson:15}. We have 
\begin{equation} 
\delta g_{tt}^{\ell=1} = \frac{2}{c^4} \B_{ab} \hat{S}^b\, x^a,  
\end{equation} 
where $\hat{S}^b := S^b/M$ is the body's spin angular momentum per
unit mass. (This expression was derived in Ref.~\cite{poisson:15} for
the specific case of a black hole, but at first post-Newtonian order it
applies equally well to any material body.) As explained in detail in
Ref.~\cite{poisson:15}, this growing term is intimately tied to the fact
that the body does not follow a geodesic in the external spacetime of
the remote objects responsible for the tidal field, but is in fact
accelerated in this spacetime; its acceleration vector --- the
acceleration of the body's local frame relative to the global,
barycentric frame ---  is given by
$-\B_{ab} \hat{S}^b/c^2$, a form of the well-known
Mathisson-Papapetrou spin force, which gives rise to the     
spin-orbit and spin-spin acceleration of a rotating body
(see Sec.~9.5 of {\it Gravity} \cite{poisson-will:14}).   

Defining an effective gravitational potential as in Appendix
\ref{app:Love}, the preceding discussion implies that
this potential possesses a dipole perturbation 
\begin{equation} 
\delta U_{\rm eff}^{\ell=1} = \frac{1}{c^2} \B_{ab} \hat{S}^b\, x^a  
\label{dUeff_dipole} 
\end{equation} 
in the body's exterior. As we have seen, this growing term is tied to
the failure of the body to move on a geodesic in the external
spacetime of the remote objects. On the other hand, the absence of a
decaying term is tied to the definition of the body's
center-of-mass. In an analogous Newtonian discussion, a multipole
expansion of the potential would normally contain a decaying, dipole
term of the form $G p_a x^a/r^3$, with $p^a = \int \rho x^a\, d^3x$
representing the mass dipole moment. But such a term is eliminated
with a judicious choice of center-of-mass, which enforces $p_a =
0$. In the relativistic setting considered here, the choice of
center-of-mass is made implicitly by demanding the absence of a
decaying term in $\delta U_{\rm eff}^{\ell=1}$.   

Inserting $S^a = I \Omega^a$ into Eq.~(\ref{dUeff_dipole}) --- refer
to Appendix \ref{app:Love} --- and incorporating the definition of
Eq.~(\ref{K_def}), we find that   
\begin{equation} 
\delta U_{\rm eff}^{\ell=1} = \frac{I}{Mc^2}\, \K_a x^a. 
\end{equation} 
Comparison with the expression of Eq.~(\ref{dU_dipole}) implies 
that $e_3 = I/(M R^2)$ in the body's exterior. We have therefore
arrived at the appropriate surface condition for the internal $e_3$.  

For the polytropic model adopted in the main text, the moment of
inertia evaluates to $I/(MR^2) = 2(\pi^2-6)/(3\pi^2) \simeq
0.26138193210$. The numerical search described in
Sec.~\ref{subsec:dipole} returned $e_3(1) \simeq 0.26138193211$ for
the same quantity. The relative numerical error is of the order of
$10^{-10}$, in line with the expectations placed on the code.   

\section{Acceleration of the center-of-mass} 
\label{app:CM} 

The dipole velocity field of Eq.~(\ref{wa_dipole}) gives rise to the
acceleration field 
\begin{equation} 
a^{\ell=1}_a := \partial_t w_a^{\ell = 1} 
= \frac{R^2}{c^2} \bigl[ 
e_1\, n_a n^b + e_5\, ( \delta_a^{\ b} - n_a n^b ) \bigr] \K_b. 
\end{equation} 
We aim to prove that this yields a vanishing acceleration for the 
body's center-of-mass,  
\begin{equation} 
a_a^{\rm CM} = \frac{1}{M} \int \rho a^{\ell=1}_a\, d^3x. 
\end{equation} 
We begin by making the substitution and carrying out the angular
integrals, using the identity $(4\pi)^{-1} \int n_a n^b\, \sin\theta
d\theta d\phi = \frac{1}{3} \delta_a^{\ b}$. This returns 
\begin{equation} 
a_a^{\rm CM} = \frac{4\pi}{3} \frac{R^2}{c^2} \K_a 
\int_0^1 \bar{\rho} (e_1 + 2 e_5) \bar{r}^2\, d\bar{r}, 
\end{equation} 
in which $\bar{\rho} := R^3\rho/M$ and $\bar{r} := r/R$. The
definition of $e_5$ in Eq.~(\ref{e5_def}) and the differential
equation (\ref{e1_diffeq}) imply that the integrand is 
\begin{equation} 
\bar{\rho} (e_1 + 2 e_5) \bar{r}^2 = \bigl( \bar{r}^3 \bar{\rho} e_1
\bigr)', 
\end{equation} 
in which a prime indicates differentiation with respect to
$\bar{r}$. Integration is immediate, and the vanishing of $\bar{\rho}$
at $\bar{r} = 1$ guarantees that 
\begin{equation} 
a_a^{\rm CM} = 0. 
\end{equation} 

\bibliography{../bib/master} 

\end{document}